\documentclass[%
 reprint,
superscriptaddress,
 amsmath,amssymb,
 aps,
]{revtex4-2}

\usepackage{xcolor}
\usepackage{graphicx}
\usepackage{dcolumn}
\usepackage{bm}
\usepackage{xr}
\makeatletter

\begin{document}

\preprint{APS/123-QED}

\title{Critical Drift in a Neuro-Inspired Adaptive Network}

\author{Silja Sormunen}
\affiliation{Department of Computer Science, Aalto University, 00076 Espoo, Finland
}%
\author{Thilo Gross}
\affiliation{
Helmholtz Institute for Functional Marine Biodiversity at the University of Oldenburg (HIFMB), 26129 Oldenburg, Germany} 
\affiliation{Alfred-Wegener Institute, Helmholtz Centre for Marine and Polar Research,  27570 Bremerhaven, Germany}
\affiliation{Institute for Chemistry and Biology of the Marine Environment (ICBM), Carl-von-Ossietzky University, 26129 Oldenburg, Germany
}
\author{Jari Saram{\"a}ki}%
\affiliation{Department of Computer Science, Aalto University, 00076 Espoo, Finland
}%

\begin{abstract}
It has been postulated that the brain operates in a self-organized critical state that brings multiple benefits, such as optimal sensitivity to input. Thus far, self-organized criticality has typically been depicted as a one-dimensional process, where one parameter is tuned to a critical value. However, the number of adjustable parameters in the brain is vast, and hence critical states can be expected to occupy a high-dimensional manifold inside a high-dimensional parameter space. Here, we show that adaptation rules inspired by homeostatic plasticity drive a neuro-inspired network to drift on a critical manifold, where the system is poised between inactivity and persistent activity. During the drift, global network parameters continue to change while the system remains at criticality.

\end{abstract}

\maketitle


{\emph{Introduction.}}---The critical brain hypothesis postulates that biological brains operate in a self-organized critical state \cite{Chialvo,Herz,pearlmutter2009hypothesis,Timme,HesseReview}. While there was initially little evidence to support this hypothesis, subsequent advances in neuroscience have made it possible to observe the characteristic power laws and avalanche dynamics associated with critical transitions, first in cell cultures \cite{Beggs2003,Yaghoubi,gireesh2008neuronal} and then in live animals and humans \cite{Kitzbichler,MeiselEcog,Hansen,Fontenele,Tagliazucchi}. Although still controversial \cite{Wilting2014years}, the critical brain hypothesis is rapidly gaining support in mainstream neuroscience, fuelled by the growing amount of experimental evidence.

This experimental evidence is complemented by a body of theory that elucidates the mechanisms that allow networks of neurons to self-organize to a critical state. Synapses that connect neurons to each other constantly self-tune their conductance through a variety of processes, collectively known as synaptic plasticity. Building on the early ideas of Bornholdt and Rohlf \cite{Bornholdt}, it has been shown with simulations that commonly observed types of synaptic plasticity, such as homeostatic and spike-time dependent plasticity, are capable of self-organizing neuronal models to a critical state \cite{MeiselModel, Droste, Levina2007, Kossio}. 

One facet of self-organized criticality that has received surprisingly little attention concerns the dimensionality of the parameter space in which the self-organization occurs. In the vast majority of studies, self-organized criticality is depicted as a one-dimensional process, where one parameter is tuned to a critical point. However, in real-world systems such as the brain, there are several and possibly very many parameters that are controlled dynamically.  
In such a high-dimensional parameter space, the states of the system that correspond to criticality can be expected to form a larger critical manifold. 

It has been conjectured that the same mechanisms that drive the system to criticality will cause a drift along the critical manifold after criticality is reached \cite{gross2021not}. While remaining critical, the system can thus continue to explore the parameter space and potentially encounter further instabilities along the way. This opens up the possibility of new phenomena such as high-codimension criticality with multiple order parameters and persistent parametric dynamics in the critical state. Understanding such phenomena may shed light on how the brain can operate in different dynamical states both sequentially and simultaneously.

In this paper, we use a simple adaptive neuro-inspired network model to show that a self-organizing system can drift on a critical manifold. This model has previously been shown to self-organize to the critical state between neuronal inactivity and persistent activity, called the onset of activity \cite{Droste}.  Here, we show that the system reaches the critical state long before the global network parameters, such as the average connectivity, reach their stable values. We carefully analyze network dynamics after the critical state has been reached, revealing the conjectured drift on the critical manifold where the ongoing plasticity continues to reshape the network structure while the system remains critical. These results provide direct evidence of the critical drift and establish an easily tractable example system where subsequent phenomena can be analyzed. 

\begin{figure*}
    \includegraphics[width=15cm]{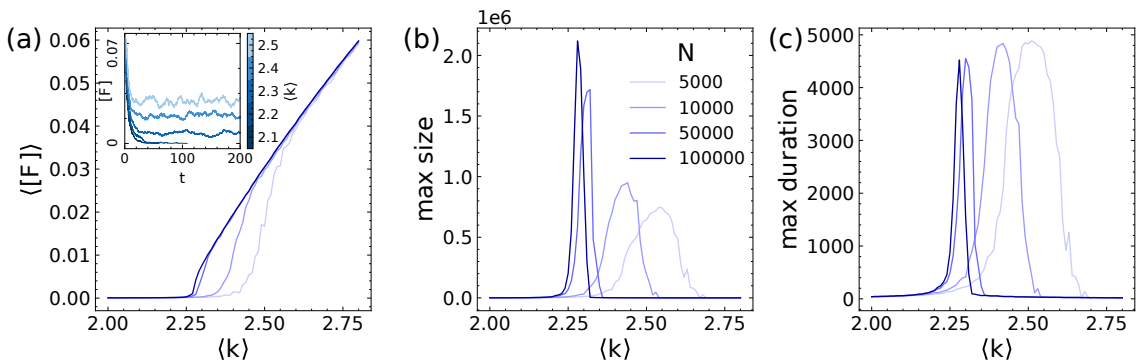}
    \caption{The onset of activity in static ER networks of different sizes $N$. (a) The average fraction of firing nodes $\langle\left[F\right]\rangle$ (average taken over time) before $t_{\rm max}=5000$, with 5\% of the nodes initialized as firing. The inset shows the time series of the fraction of firing nodes for networks of size $N=10^5$ with different mean degrees. (b-c) The maximum size and duration of finite avalanches (lasting less than $t_{\rm max}$) in 1000 successive runs. Both quantities display a sharp peak at a critical value  $\langle k \rangle^*_{\rm N, \ static}$, which moves closer to the theoretical estimate as $N$ increases. All results are averaged over 30 network realizations for each mean degree. In these and subsequent figures, we set $\beta=0.7,\delta=0.95$ and $\gamma=0.4$.}
    \label{fig:static_divergence}
\end{figure*}

{\emph{The model.}}---We investigate criticality in the model of Droste et al.~\cite{Droste} that combines stochastic neuro-inspired dynamics with adaptive network evolution. As the starting point for the adaptation, we consider directed random (Erd\H{o}s-R\'{e}nyi, ER) networks of $N$ excitable nodes and $M$ directed links with a mean degree of $\langle k \rangle = M/N $. Each node can take three discrete states: firing ($F$), refractory ($R$), or inactive ($I$). Nodes in the firing state activate their inactive neighbors stochastically at rate $\beta$ and then enter a refractory period at rate $\delta$ before transitioning back to state $I$ at rate $\gamma$. The network topology evolves on a timescale slower than the node dynamics, following rules inspired by homeostatic plasticity that strives to keep the mean firing rate of each neuron constant over the long term (see e.g.~\cite{Turrigiano}). We use a discrete update rule where firing nodes lose incoming links at rate $l $, while new links are created between random nodes at rate $g$. During the network evolution, we allow inactive nodes to fire spontaneously at rate $\eta$ to counteract activity dying out due to finite-size effects. The network dynamics and topology are evolved using the Gillespie algorithm \cite{gillespie}. The implementation is available in GitHub \cite{gitcode}.  

{\emph{Criticality in static ER networks.}}---Let us first characterize the transition from inactivity to persistent activity when the adaptation rules are switched off and no spontaneous activity is allowed. This transition separates the phase where any initialized activity dies out exponentially from the phase where exciting a random node leads to sustained activity. The average activity $\langle\left[F\right]\rangle$ acts as the order parameter of the transition (see Fig.\ 1a). In static ER networks, the mean degree $\langle k \rangle$ is the control parameter determining the overall excitability. As the firing dynamics is similar to the SIS model, the transition is expected to belong to the directed percolation universality class \cite{Hinrichsen}. In this universality class, two correlation lengths, $\xi_\parallel$ and $\xi_\perp$, diverge at the transition, with the former corresponding to the temporal dimension and the latter to the spatial (network) dimension. 

To verify that the system undergoes a continuous phase transition at a critical value $\langle k \rangle^*_{\rm static}$, we initialize several successive cascades of activity in ER networks with different mean degrees $\langle k \rangle$. These avalanches are initialized by activating one random node at a time. We then record the duration and size of the resulting avalanche, where the size indicates the number of firing events (note that one node can fire several times). We set a maximum time limit $t_{\rm max}$ so that avalanches that die out before this limit are considered finite. Their maximum size and duration are then expected to sharply peak at the critical value $\langle k \rangle^*_{\rm static}$ as a result of the diverging correlation lengths.

We observe that, as expected, the system shows the hallmarks of a continuous phase transition at a critical value  $\langle k\rangle^*_{N, \ \rm static}$, with the transition becoming sharper as $N$ increases. At this threshold, the average activity becomes non-zero and the maximum size and duration of finite avalanches diverge (Fig.~\ref{fig:static_divergence}). In line with this, the probability distributions of finite avalanche size and duration appear exponential when  $\langle k \rangle$ lies clearly under or above the critical threshold, while close to the critical value the distributions look like power laws with exponents matching the theoretical predictions for critical SIS-like systems derived in \cite{larremore} (see SI V \cite{supplementary}). The critical mean degree  $\langle k\rangle^*_{N, \ \rm static}$ depends on the transition rates $\beta,\delta$ and $\gamma$, and its value for infinite systems can be approximated with Eq.~7 from \cite{Droste}. For the parameters used here, Eq.~7 yields $\langle \hat{k} \rangle^*_{\rm static} = 2.21$, which lies slightly below the experimentally extrapolated value $\langle k \rangle^*_{N \to \infty, \ \rm static}$ (see SI I \cite{supplementary}).

\begin{figure*}
    \includegraphics[width=17cm]{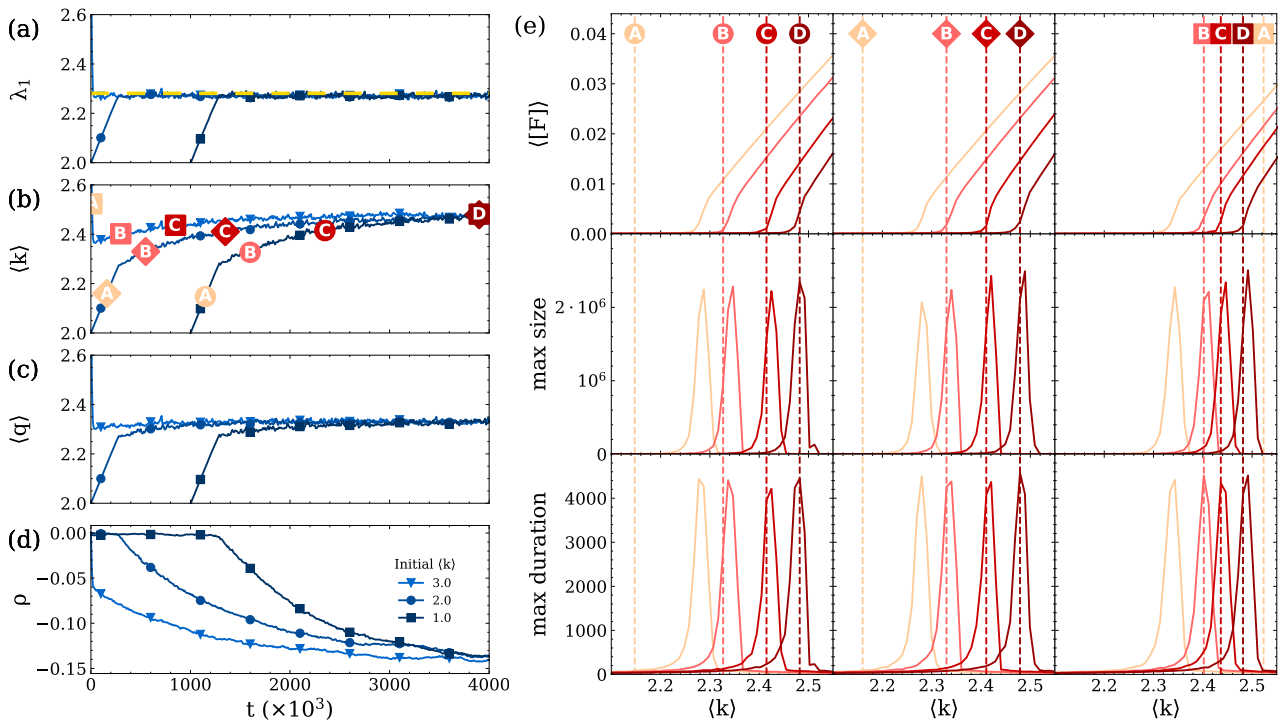}
    \caption{Critical drift in evolved networks with different initial mean degrees. (a-d) Time evolution of the  leading eigenvalue $\lambda_1$, mean degree $\langle k \rangle$, mean excess degree $\langle q \rangle$ and Pearson correlation coefficient $\rho$ of the nodes' in- and out-degrees. The simulation is initialized with $5\%$ of the nodes in the firing state. The parameter values are $N=10^5, l=10^{-3}$, $g=10^{-6}$ and $\eta = 1/(100N)$. For analysis on their effect on the drift, see SI II and III \cite{supplementary}. The yellow dashed line in (a) shows $\langle k \rangle^*_{N=10^5,\ \rm static}$ (determined in SI I \cite{supplementary}). (e) We freeze the networks at different time points shown in panel (b). These are chosen so that for each network, point A lies before the start of the drift while points B-D correspond to the drift phase. For each time point, the dashed vertical line marks the mean degree in the evolved network at that point, while the curve in the same color shows results for the networks obtained by manipulating the mean degree. The top row shows average activity $\langle [ F ] \rangle$ before $t_{\rm max}=5000$ with 5\% of the nodes initialized as firing. The other two rows show the maximum size and duration of finite avalanches in 1000 successive runs. Results are averaged over 30 network realizations for all networks obtained by manipulating $\langle k \rangle$.  We observe that during the initial phase (point A), the mean degree lies clearly under or above the onset of activity, while during the drift phase (points B-D), the networks reside at the onset of activity. As the network evolves, the onset of activity happens at higher values of $\langle k \rangle$, which confirms the existence of the manifold. Note that the divergence peaks of avalanche sizes and durations are not exactly at the point where $\langle [ F ] \rangle$ becomes non-zero (top row); however, these three measures converge to the critical value (or slightly above it due to finite values of $g$ and $l$) as $N$ is increased (see SI IV \cite{supplementary}).}  
    \label{fig:drift}
\end{figure*}

{\emph{Evidence for drift on the critical manifold.}}---Next, we switch on the plasticity rules and observe how the simulated networks evolve in time, using  ER networks of different mean degrees in the vicinity of the critical value $\langle \hat{k} \rangle^*_{\rm static} = 2.21$ as the initial condition. We follow the  evolution of the networks' key characteristics: leading eigenvalue $\lambda_1$ of the adjacency matrix, mean degree $\langle k \rangle$, mean excess degree $ \langle q \rangle$, and the Pearson correlation coefficient $\rho$ of the nodes' in- and out-degrees (Fig.~\ref{fig:drift}, a-d). 

We start by analyzing the time evolution of the leading eigenvalue $\lambda_1$. This eigenvalue reflects the overall excitability of the network, and the onset of activity is known to occur at a critical value $\lambda_1^*$ in locally tree-like networks. This has been shown previously assuming that the states of neighboring nodes are independent (see e.g. \cite{prakash2011}); here, we derive a more accurate estimate for $\lambda_1^*$ by relaxing this assumption. Using the so-called pair approximation (see SI VII \cite{supplementary}), we obtain
\begin{equation}
    \hat{\lambda}_1^* = \frac{\delta}{\beta} + \frac{\delta + \gamma/2}{\delta + \gamma},\label{eq:lambda}
\end{equation}
which is identical to $\langle \hat{k} \rangle^*_{\rm static}$ derived for static ER networks in Ref.~\cite{Droste}. Note that in general, for static ER networks, $\langle k \rangle$ and $\lambda_1$ are approximately equal. If the network structure is less random, $\langle k \rangle$ becomes a poorer approximation for the excitability. The leading eigenvalue, however, remains a more reliable indicator of  excitability, unless the network has significant degree correlations \cite{chen} or is highly structured \cite{givan}.

We observe that as the network evolves, the leading eigenvalue $\lambda_1$ reaches a stable value after a short transient (Fig.~\ref{fig:drift}a). This value lies close to $\langle k \rangle^*_{N=10^5, \ \rm static}$ and moves closer to the theoretical estimate $\hat{\lambda}_1^* = \langle \hat{k} \rangle^*_{\rm static}$ as $N$ increases (see SI I \cite{supplementary}), indicating that the system resides at criticality. 

To illustrate the drift on the critical manifold, we next analyze the evolution of
the mean degree $\langle k \rangle$. We see that $\langle k \rangle$ first changes rapidly, 
but once $\lambda_1$ stabilizes, the average rate of change in $\langle k \rangle$ decreases considerably (Fig.~\ref{fig:drift}b). Subsequently, the mean degree increases gradually and unevenly and finally settles to fluctuate around a constant value that is clearly above $\langle k \rangle^*_{N, \ \rm static}$. We interpret these qualitatively different stages as an initial phase where the dynamics approaches criticality, followed by a drift phase, where the system slides along the critical manifold. During the drift, $\lambda_1$ stays constant while $\langle k \rangle$ as well as the mean excess degree and the correlation coefficient $\rho$ (Fig.~\ref{fig:drift}c-d) keep changing.

We note that the observation of the final value of $\langle k \rangle $ differing from $\lambda_1$ is not novel per se; this has already been established for a SIRS-like system evolved with short-term homeostatic plasticity in \cite{campos}. Our novel result is the observation of the phase 
where the system is already critical before the network parameters have reached stable values.
It is also crucial to note that while $\lambda_1$ stays constant on the manifold in locally tree-like networks,  its value can change during the critical drift in networks with loops (see SI VIII \cite{supplementary}).

To confirm that the system remains critical during the drift, we 
directly assess the distance to criticality at different points in time during the network evolution. For this purpose, we initially evolve the network topology for time $t$. After this, we switch the plasticity rules off and create several replicas of the system in which we add or remove a small number of links at random. We then analyze the effect of this perturbation of the number of links on the network dynamics by examining the divergence of the size and duration of the largest finite avalanches (Fig.~\ref{fig:drift}e). During the initial phase, a large perturbation is needed to bring the system to criticality (dashed lines marked with A in Fig.~\ref{fig:drift}e). During the drift phase, however, the evolved networks reside at the divergence peak at the onset of activity (dashed lines marked with B-D). 

Furthermore, we observe that the onset of activity occurs at higher values of the mean degree as the network evolves (see SI~VI \cite{supplementary} for further illustration). In other words, 
$\langle k \rangle^*$ drifts towards higher values as the network evolves. At the same time, the network remains at criticality, as also seen in the PDFs of the sizes and durations of finite avalanches that remain unchanged during the drift and agree well with the theoretical predictions for critical SIS-like systems derived in \cite{larremore} (SI V \cite{supplementary}).

To understand why the mean degree $\langle k \rangle$ increases during the drift, we turn to analyze the characteristics of the links that the plasticity mechanism removes. As the mechanism removes links from firing nodes, links that often forward activation are likely to be erased. Intuitively, removing such links tends to reduce the overall excitability more than adding random links increases it on average. Consequently, more links need to be added than removed to keep the excitability at a constant level. This imbalance leads to $\langle k \rangle$ increasing until the most active links have been removed and the average effect of a random addition and a targeted removal even out.

This intuition can be expressed in more formal terms using the leading eigenvalue $\lambda_1$ and the corresponding left principal eigenvector. In SIS-like models \cite{Hinrichsen}, a node's eigenvector centrality (given by the left principal eigenvector) correlates with its probability of being in the firing state, and this relation is particularly strong if the system is close to criticality (see SI~IX \cite{supplementary}). Consequently, the plasticity mechanism tends to reduce the in-degrees of nodes with high centrality. As these links contribute to the magnitude of $\lambda_1$ more than a randomly chosen link on average, the removals needs to be compensated by adding links to the network to keep $\lambda_1$ close to the critical value. As time passes, the effects of link addition and removal gradually even out (see SI X \cite{supplementary}). Consequently, $\langle k \rangle$ increases more and more slowly and eventually levels off. 
This drift can be observed for a wide range of values of 
$\beta$, $\delta$, and $\gamma$, as long as the critical value of $\lambda_1$ is low enough (see SI XI \cite{supplementary}).

The leading eigenvalue depends on many topological characteristics, such as the mean excess degree $\langle q \rangle$ and the cyclic patterns in the network. In directed networks, $\langle q \rangle$ is defined as the average out-degree of nodes reached by following a link, $\langle q \rangle = \frac{1}{|\{s_{ij}\}|} \sum_{\{s_{ij}\}} k_{out,j}$, where $\{s_{ij}\}$ denotes the set of all links and $k_{out,j}$ denotes the out-degree of node $j$. It is relevant in the context of activity spreading as it
equals the expected number of new nodes that an arriving avalanche can excite.
In the considered sparse ER networks, $\langle q \rangle$ increases only slightly during the drift (Fig.~\ref{fig:drift}c), which indicates that the plasticity mechanism controls excitation mainly through restricting its growth. This is because firing nodes are likely to have predecessors with higher-than-average in-degrees, and hence the plasticity mechanism effectively reduces the out-degrees of nodes with many incoming links. This trend is reflected in the decreasing Pearson correlation coefficient $\rho$ of nodes' in- and out-degrees (Fig.~\ref{fig:drift}d) and aligns with the results in \cite{campos}, where a negative correlation between incoming and outgoing synaptic weights was found to explain the deviation of  the self-organized stable value of the branching ratio (equivalent to $\langle k \rangle$ in our model) from the mean-field prediction.

While the mean excess degree increases only slightly during the drift, it increases nonetheless. This  implies that, similarly to $\langle k \rangle$, its critical value depends on other network parameters, such as the number and configuration of cycles. Consequently, the magnitude of the increase depends largely on the original network topology. 

In this study, we have shown that rules resembling homeostatic plasticity drive simple neuro-inspired networks to drift along a critical manifold. During this drift, the network stays at the onset of activity while global network parameters continue to change. Our findings underscore that criticality should not be understood as a one-dimensional point   but rather as a high-dimensional manifold embedded in a vast parameter space, as hypothesized in \cite{Droste}. As a consequence, residing at the onset of activity does not set strict constraints to any specific network parameter, as the change in one parameter can be compensated by adjusting some other variable accordingly. This flexibility allows for considerable variation in network topology while at criticality. We emphasize that the core message of our work lies in establishing that a self-organizing system can drift along or close to a critical manifold; whether the system is exactly critical or slightly sub- \cite{priesemann2014,wilting2019} or supercritical \cite{arcangelis} likely depends on the self-organizing mechanism in question. While the model studied in this work is inspired by neuronal networks, it is very far from being biologically realistic; investigating more detailed
and realistic models is best left for future
work that builds on the foundations established here.

In the sparse random networks considered in this study, the values of the tracked parameters eventually stabilized. In real systems, however, external stimuli and a number of different driving processes continue to perturb the system. Introducing additional driving processes -- such as another type of plasticity rule \cite{girardi} -- could cause the network to continue to drift along the manifold or possibly even induce periodic parameter dynamics. If the changes in  network configuration entail changes in the dynamical behavior, the system can explore different dynamical regimes while remaining critical at all times. An interesting question concerns whether critical manifolds associated with different phase transitions intersect.
For example, can a system drift to the onset of synchrony while still remaining at the onset of activity? Exploring the structure, dynamical regimes, and intersections of these critical manifolds is an exciting avenue for future research.

{\emph{Acknowledgments.}}---We acknowledge the computational resources provided by the Aalto University Science-IT project.

\InputIfFileExists{bibliography.bbl}


\end{document}



\title{Critical Drift in a Neuro-Inspired Adaptive Network\\ \large Supplemental material}

\author{Silja Sormunen}
\affiliation{\mbox{Department of Computer Science, Aalto University, 00076 Espoo, Finland}
}%
\author{Thilo Gross}
 
\affiliation{
Helmholtz Institute for Functional Marine Biodiversity at the University of Oldenburg (HIFMB),\mbox{26129 Oldenburg, Germany}}
\affiliation{\mbox{Alfred-Wegener Institute, Helmholtz Centre for Marine and Polar Research, 27570 Bremerhaven, Germany}}
\affiliation{Institute for Chemistry and Biology of the Marine Environment (ICBM), \mbox{Carl-von-Ossietzky University, 26129 Oldenburg, Germany}
}
\author{Jari Saram{\"a}ki}%
\affiliation{\mbox{Department of Computer Science, Aalto University, 00076 Espoo, Finland}
}%

\maketitle

\renewcommand\thesubsection{\Roman{subsection}}

\renewcommand\thesubsubsection{\Alph{subsubsection}}

\subsection{Critical values of $\langle k \rangle$ and $\lambda_1$}

\begin{figure}[h!]
    \includegraphics[width=8.5cm]{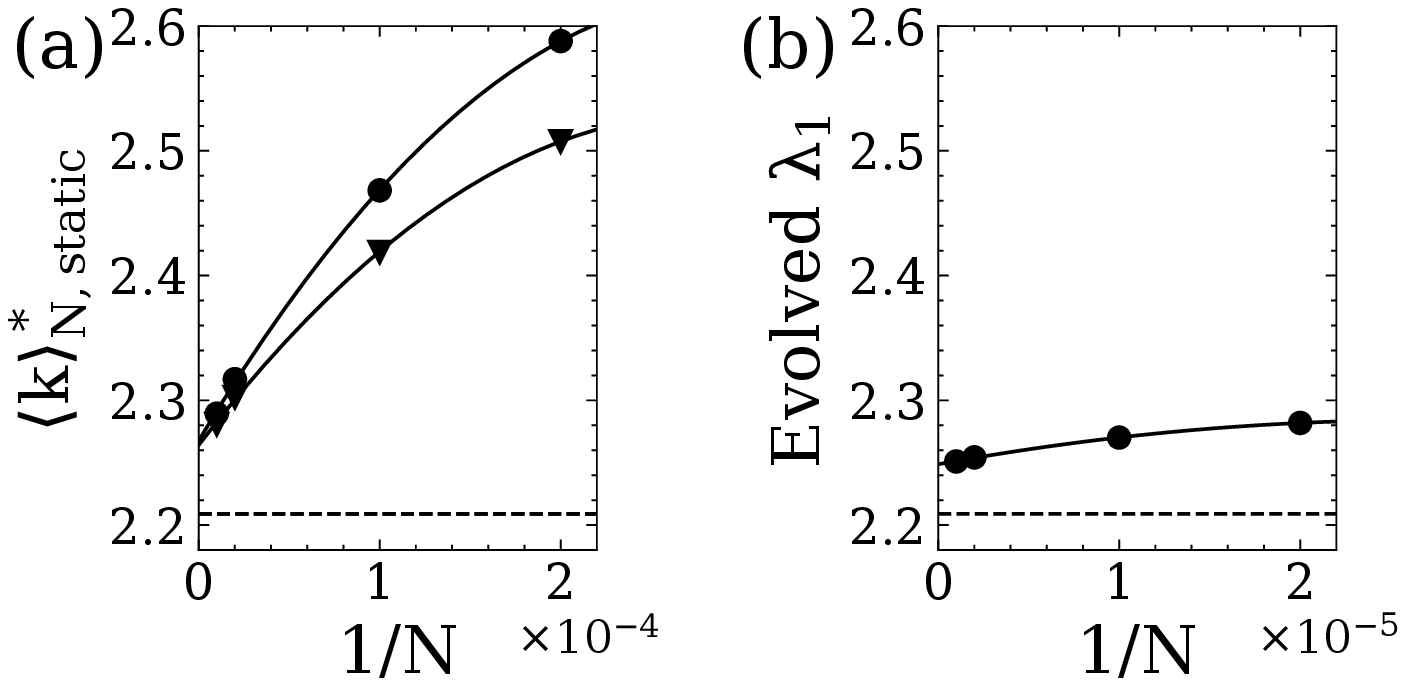}
    \caption{(a) Critical value $\langle k \rangle^*_{N,\ \rm static} $ in static ER networks as a function of $1/N$ calculated using avalanche sizes (circles) and durations (triangles). The markers overlay the 95\% confidence intervals. The dashed line marks the theoretical estimate $ \langle \hat{k} \rangle^*_{\rm static} = 2.21$. (b) The mean value of $\lambda_{1,N}$ in evolved networks during the drift phase as a function of $1/N$. The initial degree is 3 and $\mu$ is set to $1/(100N)$. The extrapolated value $\lambda^*_{1,N \to \infty}$ lies close to the value $\langle k \rangle^*_{N \to \infty,\ \rm static}$. }
    \label{fig:1/Nfig}
\end{figure}

Fig \ref{fig:1/Nfig}(a) shows the empirical estimates for the critical mean degree $\langle k \rangle^*_{N,\ \rm static} $ in static ER networks as a function of $1/N$.
The critical values are determined by first finding the mean degree for which the size/duration of finite avalanches in 1000 successive runs is maximal, then repeating this procedure for 30 network realizations and taking the average of the found values. We fit a quadratic polynomial to the points to find the intersection with the $y$-axis. This intersection lies around $\langle k \rangle^*_{N \to \infty, \ \rm static} = 2.27$, slightly above the theoretical estimate $ \langle \hat{k} \rangle^*_{\rm static} = 2.21$.

In the evolved ER networks of size $N=10^5$, the leading eigenvalue $\lambda_1$ self-organizes to a value close to $\langle k \rangle^*_{N=10^5, \ \rm static}$ during the drift phase, and this value moves closer to the theoretical estimate $\hat{\lambda}_1^* = \langle \hat{k} \rangle^*_{\rm static}$ as the network size increases [Fig.~\ref{fig:1/Nfig}(b)].

\subsection{Effect of network size on the drift}
\label{appendix:size}

\begin{figure}[h!]
    \includegraphics[width=5cm]{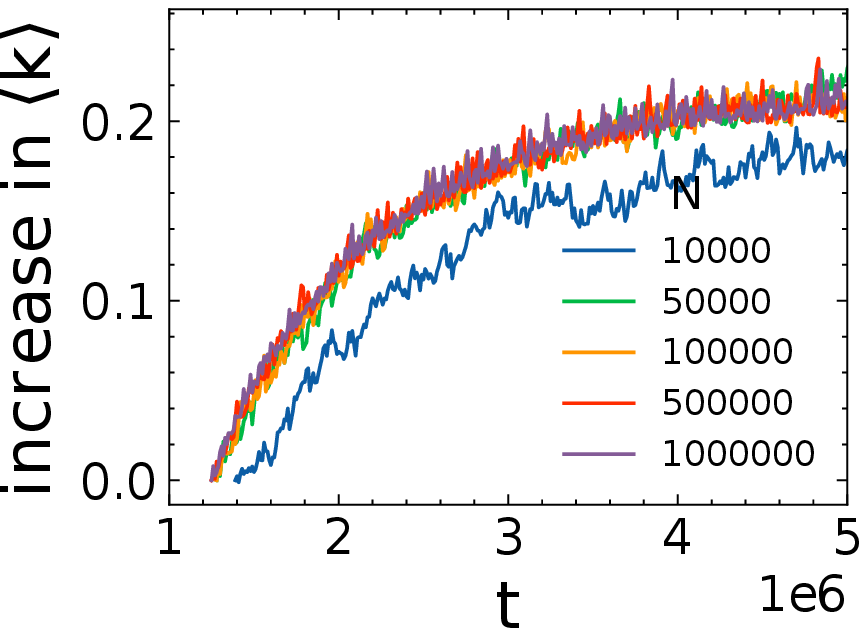}
    \caption{Increase in $\langle k \rangle$ during  the drift phase. Initial mean degree is 1 and the spontaneous firing rate equals $\frac{1}{100N}$.}
    \label{fig:sizefig}
\end{figure}

We verify that the increase in the mean degree $\langle k \rangle$ during the drift phase does not arise from some finite size effect. To this end, we first calculate the average value of $\lambda_1$ after it has clearly stabilized, and subsequently define the drift to start when $\lambda_1$ first crosses this value. Next, we subtract the value of $\langle k \rangle$ at the start of the drift from the subsequent values of $\langle k \rangle$ during the drift.  After this normalization, it becomes evident that increasing $N$ has very little effect on the drift for networks larger than $N=50000$ [Fig.~\ref{fig:sizefig}].

\subsection{Parameter considerations} \label{appendix:parameters}

\begin{figure}[h!]
    \includegraphics[width=10cm]{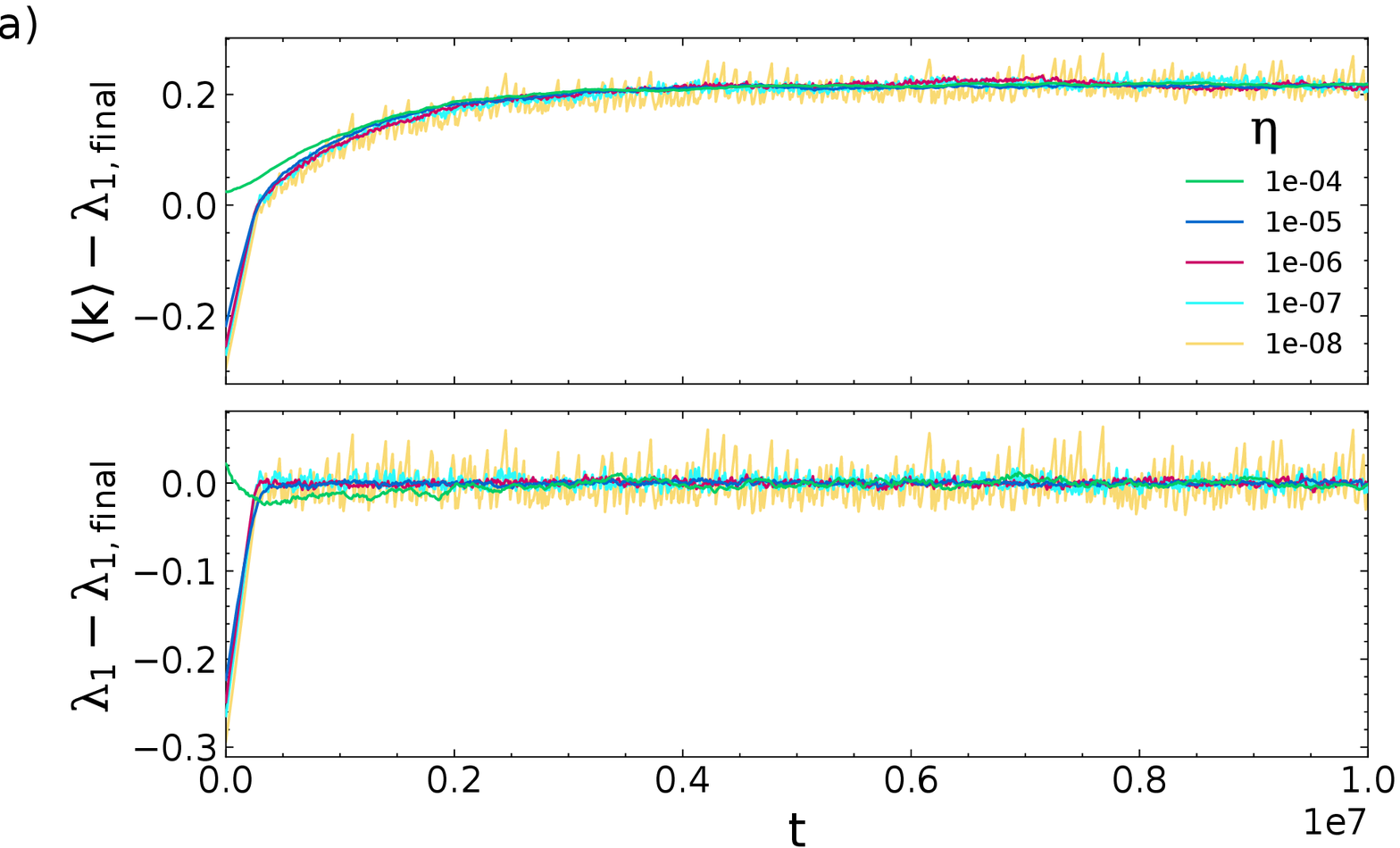}
    \includegraphics[width=10cm]{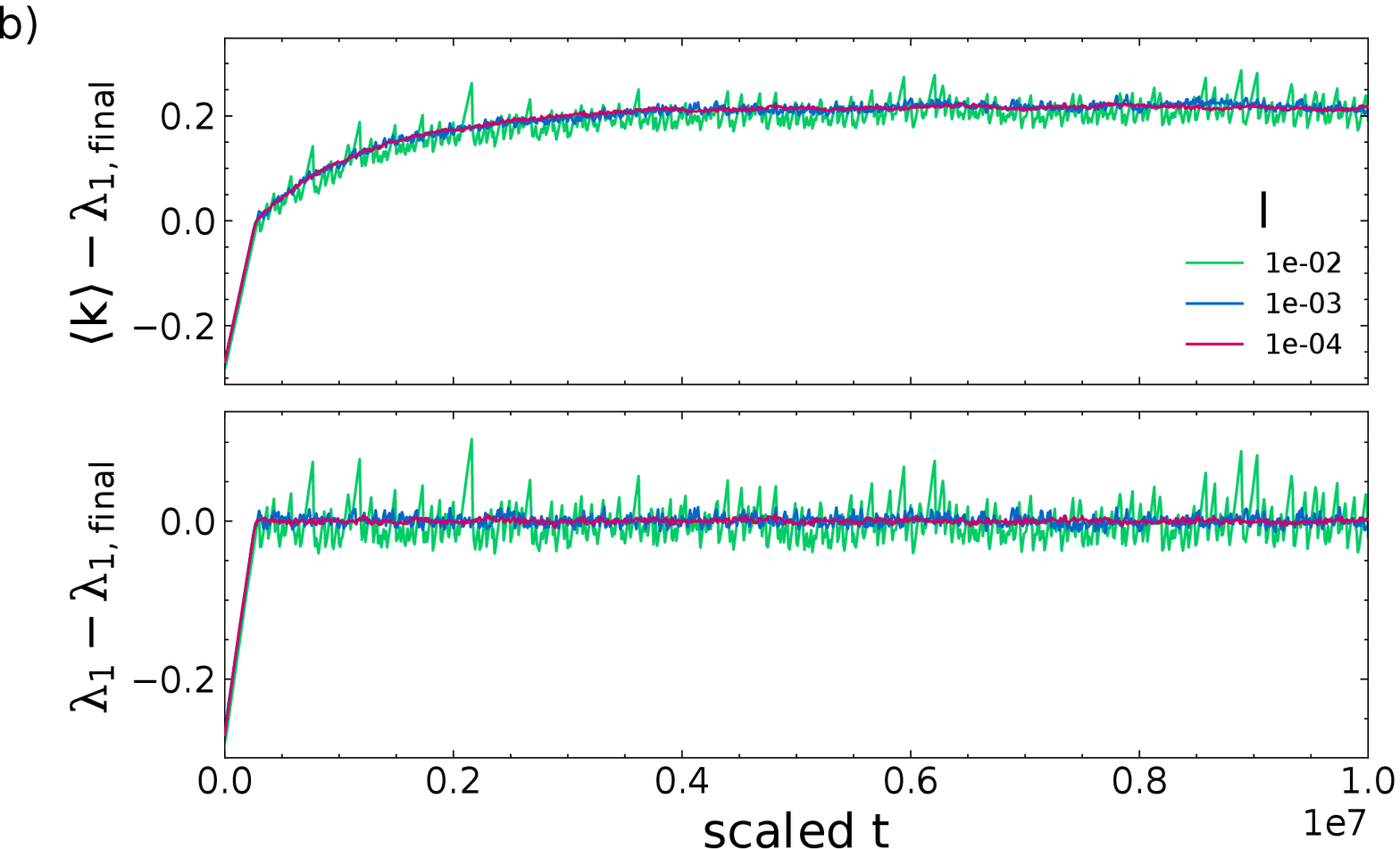}
    \includegraphics[width=10cm]{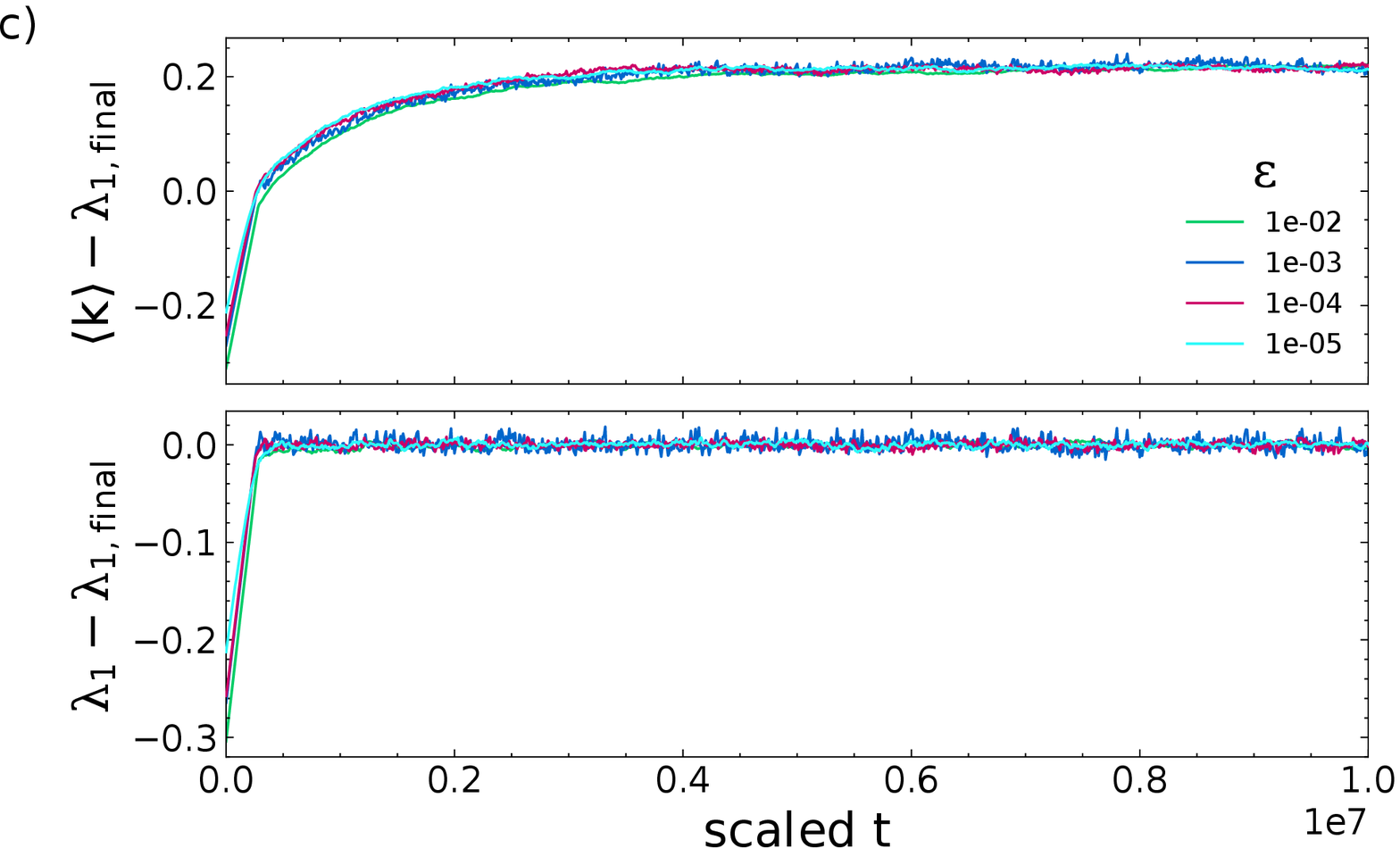}
    \caption{Time evolution of $\langle k \rangle$ and $\lambda_1$ normalized by the stabilized value of $\lambda_1$ in a network of size $N=10^5$ for different parameter values of $\eta$, $l$ and $\epsilon$. The default parameter values are $\eta=10^{-7}, l=0.001$ and $\epsilon=0.001$. As the parameters $l$ and $\epsilon$ control the timescale of the drift, we have additionally scaled the time series in figures (b) and (c) to correspond to the pace of the drift for the default values. The scaling is done by multiplying the $x$-axis by the factor by which the parameter value is larger than 0.001.}
    \label{fig:param_fig}
\end{figure}

\subsubsection{Spontaneous firing rate $\eta$}
As discussed in \cite{Droste}, the spontaneous firing rate $\eta$ plays a role in how close to the critical threshold $\lambda_{1,N}^*$ the network self-organizes to; when $\eta$ is increased, the stabilized value of $\lambda_{1,N}$ decreases. This happens because spontaneous activity contributes to the overall level of activity, which the self-organization process strives to keep at a constant level. However, as demonstrated in \cite{Droste}, the effect of the exact value of $\eta$ on the distance to criticality becomes negligible for small enough $\eta$ and large enough networks. We observe that for a network of size $N=10^5$, the value of $\eta$ has little effect also on the drift of $\langle k \rangle$ when $\eta < 10^{-4}$ [Fig. \ref{fig:param_fig}(a)].

In networks of finite size, the value of $\eta$ affects how smoothly the mean degree changes. If $\eta$ is low, intervals between subsequent bursts of activity lengthen, and the network is likely to cross well into the supercritical regime before a new avalanche takes place and the link removal mechanism is activated. Consequently, the fluctuations between sub- and supercritical states become more pronounced. However, this effect subsides as the network size increases.

\subsubsection{Network evolution parameters}

In general, self-organized criticality is possible only if the system's dynamics can be separated into two parts; the dynamics that becomes critical, and the controlling dynamics that steers the former to criticality. Dynamically speaking, this division is justified only if a timescale separation exists between the two. Consequently, the firing dynamics need to have a faster timescale than the network evolution, \emph{i.e.} $\beta,\delta,\gamma \gg l,g$. In addition, 
the rate $l$ at which nodes lose links needs to be clearly higher than the rate $g$ at which they gain links. As we are interested in the magnitude of $g$ compared to $l$, we define $g=\epsilon l$, where $\epsilon$ is a scaling parameter. This second time-scale separation stems from the fact that the plasticity rules act locally instead of controlling the average activity in a more centralized manner. Since firing nodes characterize only the active phase but inactive nodes are common in both the quiescent and the active phase, local excitability has to be increased gradually for inactive nodes and decreased quickly for active nodes. As shown in \cite{Droste}, the network self-organizes to criticality in the limit   $l \to 0, \epsilon \to 0$.

The exact values of $l$ and $\epsilon$ have little effect on the magnitude of increase in $\langle k \rangle$ during the drift phase [see Fig.~\ref{fig:param_fig}(b),(c)]. Both parameters control the timescale of the drift; if one of the parameters is decreased by an order of magnitude, the duration of the drift correspondingly increases by an order of magnitude.

\subsection{Scaling of signatures of criticality}
\label{appendix:signatures}

In Figure 2(e) in the main text, we observed that the divergence peaks of avalanche sizes and durations did not always align with the point where the average activity $\langle [F] \rangle$ abruptly increased. Here, we show that this discrepancy is a finite-size effect, and that the different ways of measuring criticality converge as the network size increases.

\begin{figure}[! h]
 \centering
    \includegraphics[width=14cm]{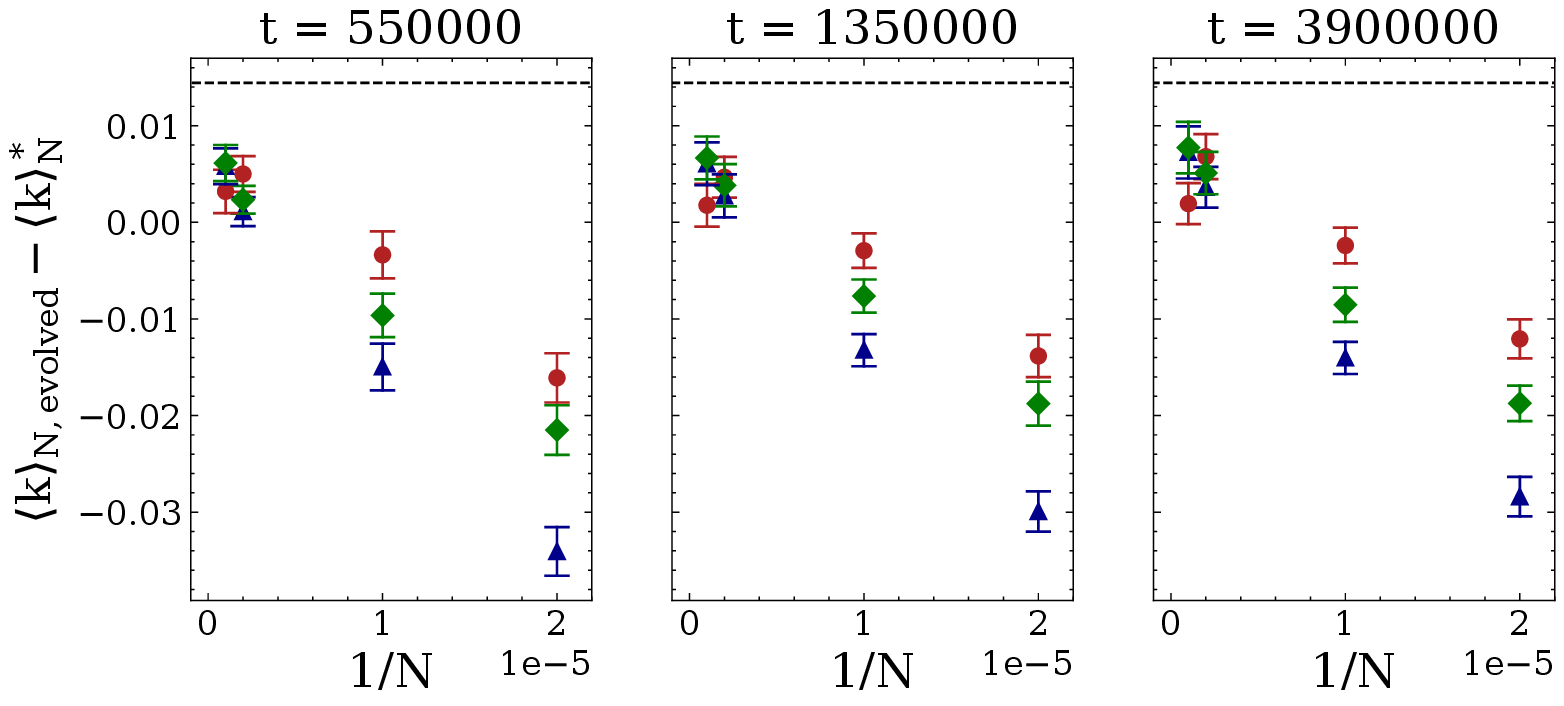}
    \caption{Distance to criticality at different times during the network evolution as a function of inverse network size $1/N$. Distance to criticality is measured based on the average activity (red circles), maximum size (blue triangles) and duration (green diamonds) of finite avalanches. The initial mean degree is 2 and the spontaneous firing rate is set to $1/(100N)$. For each $N$ and each timepoint, we first determine the estimates of $\langle k \rangle ^*_{N}$ for one time series by averaging over 10 rounds of degree manipulations. The figures display the average and the 95\% confidence intervals over 30 time series.   
    Note that the resolution of the $y$-axis is 0.01 before any averaging.}
    \label{fig:signatures}
\end{figure}

To show this, we repeat the procedure of Fig.~2(e) in the main text for networks of different sizes.  We then determine the critical degree $\langle k \rangle ^*_{N}$ in three different ways; either based on where the curve for maximum size or duration of finite avalanches reaches its maximal value, or by looking at where the average activity $\langle [F] \rangle$ changes the most between two consecutive mean degrees. 
We observe that these different ways to determine $\langle k \rangle ^*_{N}$ converge as the network size $N$ increases [Fig.~\ref{fig:signatures}].

Looking at Fig.~\ref{fig:signatures}, we observe that as $N$ increases, the mean degree of the evolved networks seems to converge to a value slightly above $\langle k \rangle ^*_{N}$. In fact, this is to be expected; as discussed in the previous section, the network self-organizes to criticality in the limit $l,\epsilon \to 0$. Hence, the evolved mean degree is expected to lie slightly above the critical value for finite values of $l$ and $\epsilon$.

\cite{Droste} derive a theoretical estimate for this deviation for specific values of $l$ and $\epsilon$ under the assumption that the network is an ER network (see Eq. (11g) of their paper). In Fig.~\ref{fig:signatures}, the dashed line marks this theoretical deviation from the theoretical critical value $\langle k \rangle^* = 2.21$. We observe that the mean degree of the simulated networks  would seem to deviate from the critical value less than suggested by Eq. (11g). This may be affected by the evolved networks violating the assumption of the network being an ER network. In addition, since the theoretical deviation depends on $\langle k \rangle ^*_{N}$ for $l, \epsilon \neq 0$, the true distance to criticality may vary slightly as $\langle k \rangle ^*_{N}$ increases during the drift. Finally, the mean degree is expected to converge to the theoretical value from below only as the spontaneous activity goes to zero.

\subsection{PDFs of avalanche sizes and durations}
\label{appendix:pdfs}

\begin{figure}[! h]
    \includegraphics[width=15cm]{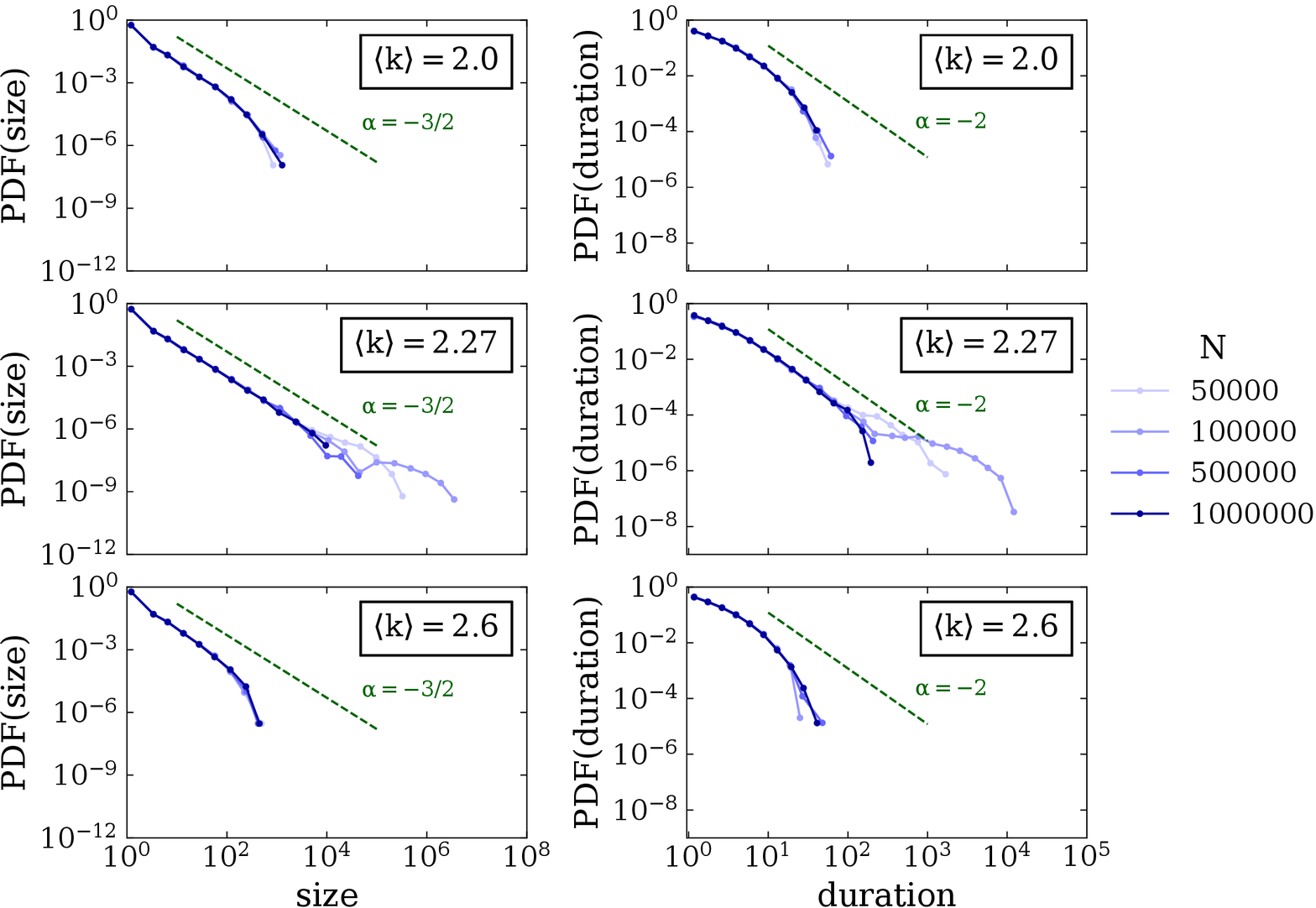}
    \caption{PDFs of the sizes and durations of finite avalanches in static random networks with different mean degrees and varying network sizes. The mean degrees of the top and bottom rows are chosen to be below and above criticality, respectively, while the middle row corresponds to the experimentally extrapolated critical value $\langle k \rangle^*_ {\rm static, N \to \infty}=2.27$. The number of runs for each mean degree is originally 10000, but only avalanches lasting less than $t_{\rm max}=50000$ are considered.   The green dashed lines mark the slopes derived for critical SIS-like systems in \cite{Larremore}. The x-axis in the right-hand column starts at $0.95$, which is the expected value of the time a node spends in the firing state before transitioning to refractory. 
    }
    \label{fig:static_sizedists}
\end{figure}

Figure \ref{fig:static_sizedists} shows the PDFs of  the sizes and durations of finite avalanches in static networks with different mean degrees. When the mean degree equals the experimentally extrapolated value $\langle k \rangle^*_ {\rm static, N \to \infty}=2.27$, the distributions resemble power law distributions with exponents matching the theoretical predictions for critical SIS-like systems derived in \cite{Larremore}. When the mean degree lies clearly below or above the critical value, the distributions' tails decay markedly faster.

We observe that the tail of avalanche sizes is in general heavier than that of avalanche durations, which aligns with the results of \cite{bak} and \cite{Larremore}. The beginning of the distribution of avalanche durations (right-hand column in Fig.~\ref{fig:static_sizedists}) is furthermore affected by the exponential distribution of the times that nodes spend in the firing state. 

\begin{figure}[! h]
 \centering
    \includegraphics[width=14cm]{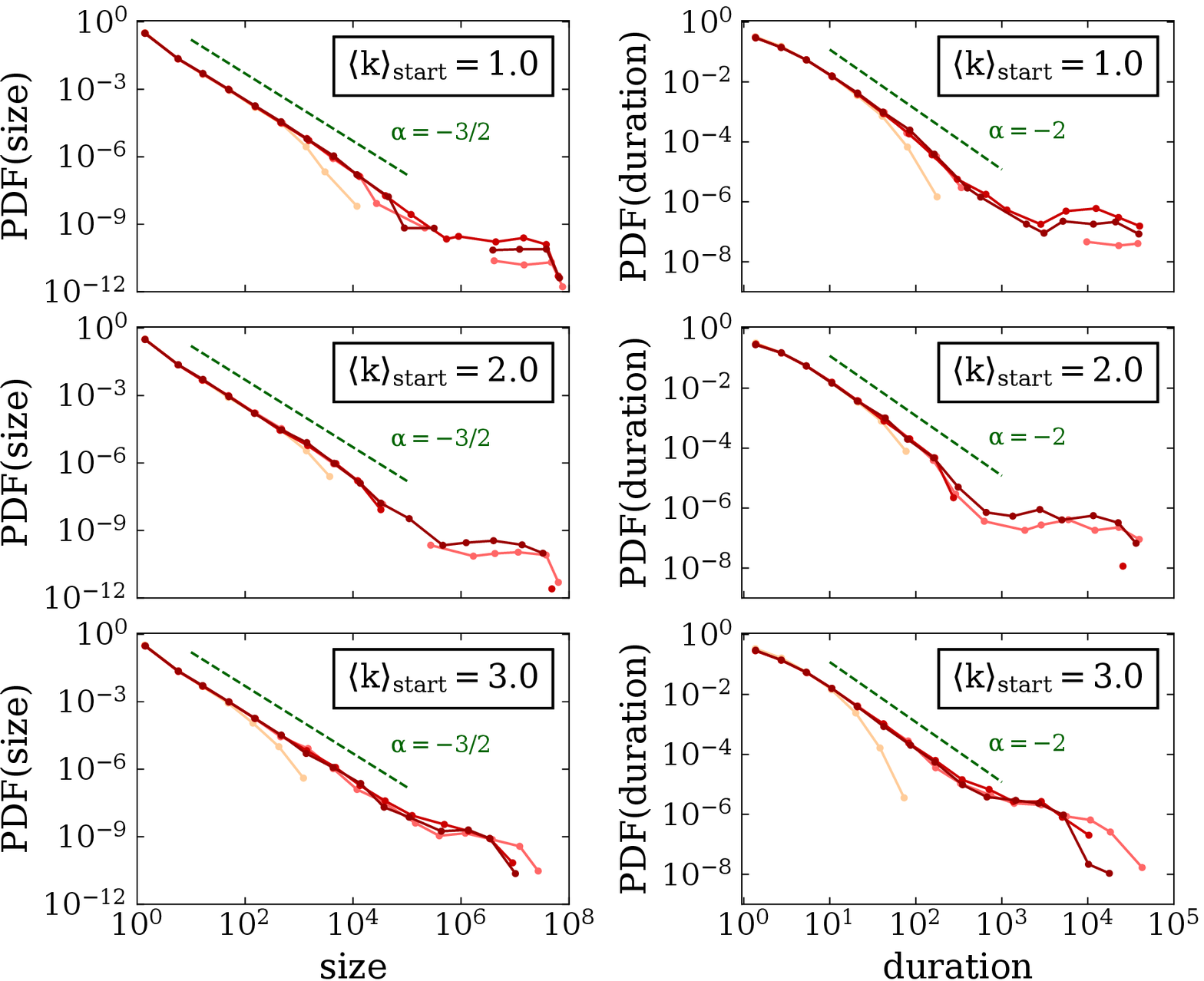}
    \caption{PDFs of the sizes and durations of finite avalanches in the evolved networks at different stages during the network evolution. The colors correspond to the distinct timepoints shown in Fig.~2 in the main text. Before the drift (light orange), the distributions decay faster, while during the drift (red shades) the distributions are heavy-tailed. The green dashed lines mark the slopes derived for critical SIS-like systems in \cite{Larremore}. The network size is $N=500000$ and the spontaneous firing rate is equal to $1/(100N)$, while all other parameters controlling the dynamics and the network evolution are identical to those used in Fig.~2 in the main text. Number of runs for each mean degree is originally 10000, but only avalanches lasting less than $t_{\rm max}=50000$ are considered.  }
    \label{fig:evolved_sizedists}
\end{figure}

\begin{figure}[! h]
 \centering
    \includegraphics[width=14cm]{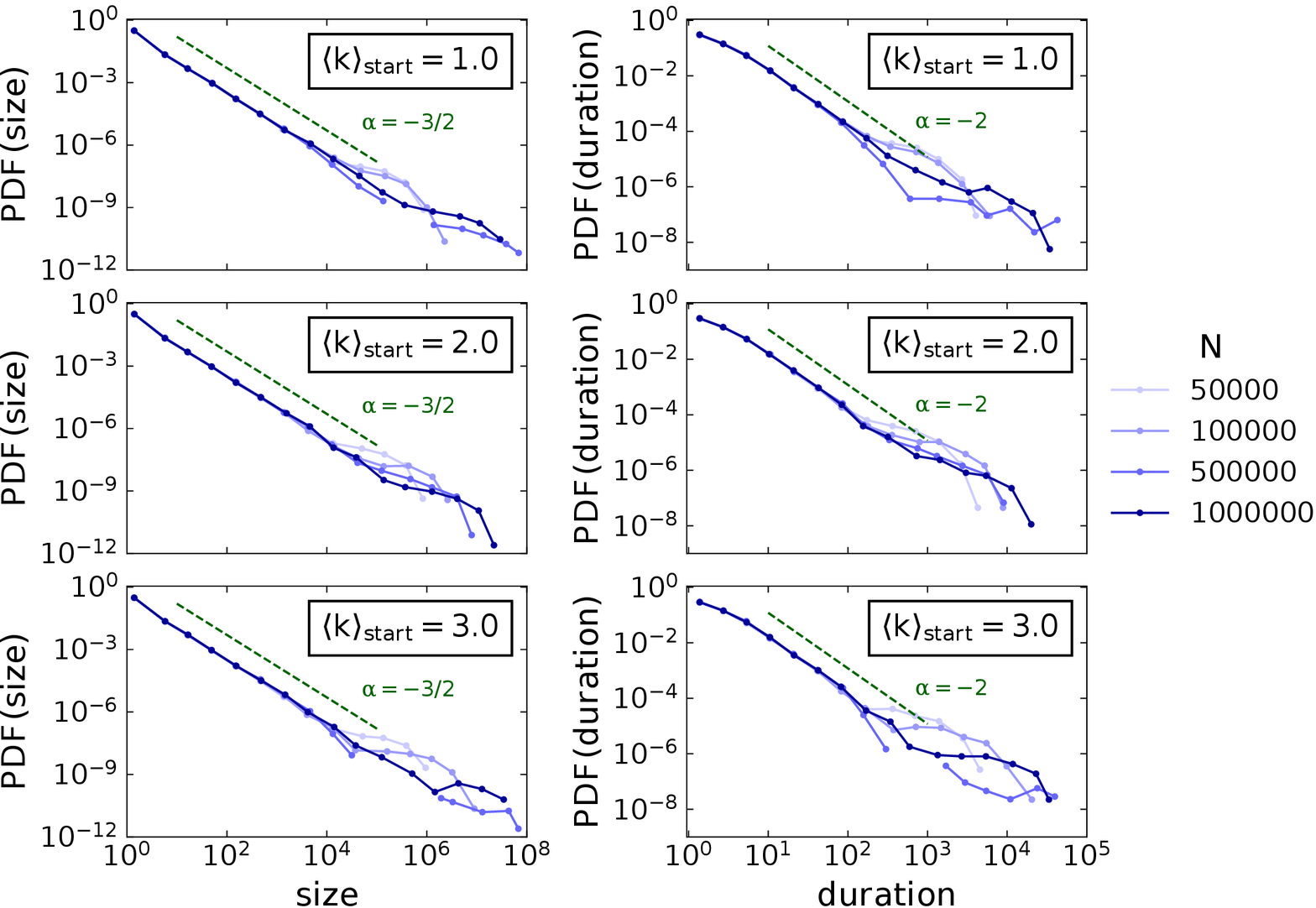}
    \caption{PDFs of evolved networks at time $t = 3900 000$ for different network sizes. The networks have been evolved with spontaneous firing rate equal to $1/(100N)$.}
    \label{fig:evolved_sizedists_Ns}
\end{figure}

To further verify that our self-organizing system resides at criticality during the drift phase, we plot the PDFs of avalanche sizes and durations at different points during the drift [Fig.~\ref{fig:evolved_sizedists}]d. We observe that these distributions resemble power law distributions with exponents matching the theoretical predictions at criticality. Most of the distributions have a small bump in their tail, which might indicate that the systems are slightly supercritical. This result would align with the theoretical prediction derived in \cite{Droste} already discussed in sections \ref{appendix:parameters} and \ref{appendix:signatures}, namely that our model is strictly critical in the limit $l,\epsilon \to \infty$ and deviates slightly to the supercritical regime for finite parameter values.
When the network size increases, this bump moves further towards the tail [see Fig. \ref{fig:evolved_sizedists_Ns}].

\clearpage

\subsection{Illustration of the critical drift} 
\label{appendix:insetfig}

\begin{figure}[h!]
    \includegraphics[width=11cm]{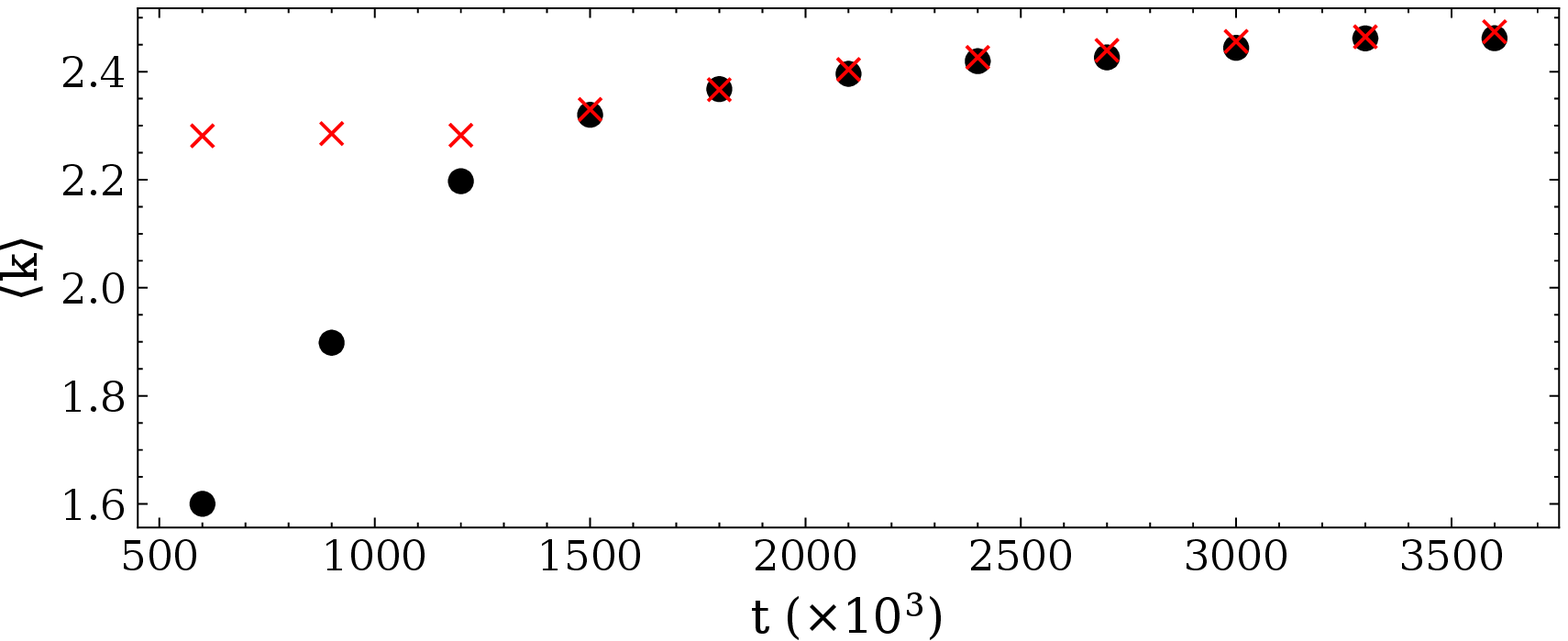}
    \caption{The mean degree $\langle k \rangle$ (black circles) and the critical $\langle k \rangle^*_{N=10^5}$ (red crosses) in a network evolved with the plasticity rules. The initial degree is 1 and the parameters are set to $N=10^5, l=10^{-3}$, $g=10^{-6}$ and $\eta = \frac{1}{100N}$. The values of $\langle k \rangle^*_{N=10^5}$ are obtained by first manipulating $\langle k \rangle$ of the evolved network as in Fig.~2(e) in the main text and then finding $\langle k \rangle^*_{N=10^5}$ based on avalanche durations as in Fig.~\ref{fig:1/Nfig} of SM. }
    \label{fig:insetfig}
\end{figure}

Figure~\ref{fig:insetfig} illustrates that the value of the critical mean degree $\langle k \rangle^*_{N=10^5}$ increases during the drift phase. In addition, we observe that during the initial phase, the mean degree of the evolved network lies below $\langle k \rangle^*_{N=10^5}$ while during the drift phase, the system resides at the onset of activity.

\subsection{Leading eigenvalue as an indicator of criticality} \label{appendix:pair}

The leading eigenvalue is known to reflect the critical threshold in epidemic models such as SIS and SIRS (identical to our static IFRI model) on undirected networks. However, previous proofs for the SIRS/IFRI model have mostly relied on the assumption that the states of two neighboring nodes are independent. This assumption leads to a threshold condition $\lambda_1 = \delta/\beta$ (see e.g. \cite{prakash2011}), which does not  accurately predict the critical threshold in our model. 

To improve the accuracy, we no longer assume that the probability of a link connecting a node in state $X$ to a node in state $Y$ is simply given by the product of the states' probabilities. Instead, we derive evolution equations for the probability of node $i$ being in state $X$ and node $j$ being in state $Y$, which we denote by $[X_i Y_j]$. Often, these probabilities depend on the probabilities of  two-link structures, say, the probability of having a $FI$-link followed by a $II$-link or two $F$-nodes pointing to the same $I$-node. To avoid rendering the system overly complex, we approximate these probabilities using the link and node probabilities. For example, given a path $k \to j \to i$, the probability of the first link being an $FI$-link and the second link being an $II$-link would be given by $[F_k I_j][I_j I_i] / I_j$, where $I_j$ denotes the probability of node $j$ being in state $I$.
This type of approximation -- called the pair approximation -- is used in \cite{Mata} to study the critical threshold for $\lambda_1$ in the undirected SIS model, and in \cite{Droste} to derive a critical value for the mean degree in the SIRS/IFRI model when the exact network structure is not known. Here, we assume that the adjacency matrix $\mathbf{A}$ is known and derive a critical threshold for the leading eigenvalue. 

With the pair approximation, we obtain the following differential equations:
\begin{align}         \dot{F}_i&= -\delta     F_{i} + \beta\sum_{j=1}^ N A_{ji}[F_j I_i] \label{startdiff}\\
    \dot{R}_i&= \delta F_{i} - \gamma R_{i} \\
    \dot{[F_j I_i]} &= A_{ji}\Big(\beta\frac{[I_j I_i]}{I_j}\sum_{k=1}^N A_{kj} [F_k I_j] + \gamma[F_j R_i]-\beta\frac{[F_j I_i]}{I_i}\sum_{k \neq j}^N A_{ki}[F_k I_i] - (\beta+\delta) [F_j I_i]\Big)\\
    \dot{[I_j I_i]} &= A_{ji}\Big(\gamma[I_j R_i] + \gamma[R_j I_i] - \beta\frac{[I_j I_i]}{I_i}\sum_{k \neq j}^N A_{ki}[F_k I_i] -\beta\frac{[I_j I_i]}{I_j}\sum_{k =1}^N A_{kj}[F_k I_j]\Big)\\
    \dot{[F_j R_i]} &= A_{ji}\Big(\delta[F_j F_i] + \beta\frac{[I_j R_i]}{I_j}\sum_{k =1}^N A_{kj}[F_k I_j]- (\gamma+\delta)[F_j R_i]\Big)\\
    \dot{[I_j R_i]} &= A_{ji}\Big(\gamma[R_j R_i] + \delta[I_j F_i]-\gamma[I_j R_i] - \beta\frac{[I_j R_i]}{I_j}\sum_{k =1}^N A_{kj}[F_k I_j]\Big)\\
    \dot{[R_j I_i]} &= A_{ji}\Big(\delta[F_j I_i] + \gamma[R_j R_i]- \beta\frac{[R_j I_i]}{I_i}\sum_{k \neq j}^N A_{ki}[F_k I_i] - \gamma[R_j I_i]\Big)\\
    \dot{[F_j F_i]} &= A_{ji}\Big(\beta\frac{[I_j F_i]}{I_j}\sum_{k =1}^N A_{kj}[F_k I_j]+ \beta\frac{[F_j I_i]}{I_i}\sum_{k \neq j}^N A_{ki}[F_k I_i] + \beta[F_j I_i] - 2\delta[F_j F_i]\Big)\\
    \dot{[R_j R_i]} &= A_{ji}\Big(\delta[F_j R_i]+\delta[R_j F_i] - 2\gamma[R_j R_i]\Big)\\
    \dot{[R_j F_i]} &=  A_{ji}\Big(\beta\frac{[R_j I_i]}{I_i}\sum_{k \neq j}^N A_{ki}[F_k I_i] + \delta[F_j F_i] - (\gamma+\delta)[R_j F_i]\Big), \label{enddiff}
\end{align}

where we use the conservation laws
\begin{align}
    I_i &= 1 - F_i - R_i\\
    [I_j F_i] &= 1 - [F_jF_i]-[F_jI_i]-[F_jR_i]-[I_jI_i]-[I_jR_i]-[R_jF_i]-[R_jI_i]-[R_jR_i].
\end{align}

The onset of activity occurs when the so-called trivial fixed point $(I_i = [I_j I_i] = 1$ for all $i,j$) becomes unstable. From dynamical systems theory we know that the fixed point loses stability when the leading eigenvalue of the associated Jacobian matrix crosses from negative to positive, \emph{i.e.} $\lambda_1(\mathbf{J}) = 0$. The Jacobian associated with equations \ref{startdiff}-\ref{enddiff} can be interpreted as a block matrix of size $(2N+8N^2)\times(2N+8N^2)$ consisting of four blocks of size $N \times N$, 16 blocks of size $N \times N^2$, 16 blocks of of size $N^2 \times N$ and 64 blocks of size $N^2 \times N^2$. At the trivial steady state, all blocks except for $\mathbf{X}$, $\mathbf{Y}$ and $\mathbf{Z}$ are diagonal matrices. The diagonal matrices corresponding to variables $F-R$ are identity matrices multiplied by the constant given below. The other diagonal matrices are given by $\mathrm{diag}(\mathrm{vec}(\mathbf{A}))$ multiplied by the constant given below:
\begin{equation}
\mathbf{J} = \begin{matrix}
 & F & R & FF & FI &FR & II &IR &RF & RI &RR \\
 F & -\delta & & & \mathbf{X}\\
 R & \delta & -\gamma\\
 FF & &  & -2\delta & \beta \\
 FI & & & & \mathbf{Y} & \gamma\\
 FR & & & \delta & & -\gamma-\delta \\
 II & & & & \mathbf{Z} & & &\gamma & & \gamma\\
 IR & & &-\delta & -\delta &-\delta&-\delta&-\delta-\gamma&-\delta&-\delta&-\delta+\gamma\\
 RF & & & \delta & & & & & -\gamma-\delta& &\\
 RI & & & & \delta & & & & & -\gamma & \gamma\\
 RR  & &  & & & \delta &  & & \delta & & -2\gamma\\
\end{matrix}
\end{equation}

To find the stability condition, it suffices to examine a smaller part of the Jacobian. To see this, we divide the Jacobian into four blocks, where $\mathbf{B_1}$ corresponds to variables $F-[FR]$ and $\mathbf{B_3}$ to $[II]-[RR]$. The eigenvalue equation is then given by
\begin{equation}
    \begin{bmatrix}
     \mathbf{B_1} & 0 \\
     \mathbf{B_2} & \mathbf{B_3}
    \end{bmatrix}
    \begin{bmatrix}
     \mathbf{e_1} \\
     \mathbf{e_2} 
    \end{bmatrix}
     =\lambda(\mathbf{J}) \begin{bmatrix}
     \mathbf{e_1} \\
     \mathbf{e_2}
    \end{bmatrix},
\end{equation}
where the eigenvector $\mathbf{e}$ has been split into two parts $\mathbf{e_1}$ and $\mathbf{e_2}$. The multiplication results in equations
\begin{align}
    \mathbf{B_1}\mathbf{e_1} &= \lambda(\mathbf{J})\mathbf{e_1} \\
    \mathbf{B_2}\mathbf{e_1} + \mathbf{B_3}\mathbf{e_2} &= \lambda(\mathbf{J})\mathbf{e_2}.
\end{align}
The first equation shows that $\lambda(\mathbf{J})$ must be an eigenvalue of $\mathbf{B_1}$ if $\mathbf{e_1}$ is not a zero vector. On the other hand, if $\mathbf{e_1}$ equals zero, $\lambda(\mathbf{J})$ must be an eigenvalue of $\mathbf{B_3}$. Consequently, the eigenvalues of  $\lambda(\mathbf{J})$ are given by the eigenvalues of $\mathbf{B_1}$ and $\mathbf{B_3}$. However, since the eigenvalues of $\mathbf{B_3}$ are non-positive constants, it suffices to examine the eigenvalues of $\mathbf{B_1}$.

With similar logic, we can divide $\mathbf{B_1}$ into four blocks where one block is a zero matrix. Consequently, the eigenvalues of $\mathbf{B_1}$ are given by the blocks formed by variables $F-R$ and $[FF]-[FR]$. The eigenvalues of the first block are non-positive constants, meaning that the stability conditions can be found by examining the matrix  
\begin{equation}
    \mathbf{C} = \begin{bmatrix}
 -2\delta & \beta & 0 \\
 0 & \mathbf{Y}  &\gamma\\
\delta & 0 & -\gamma-\delta
\end{bmatrix},
\end{equation}
where 
$\mathbf{Y}  = \beta \mathbf{M} - (\beta+ \delta)\mathrm{diag}(\mathrm{vec}(\mathbf{A}))$ and $\mathbf{M}$ is of the form

\begin{equation} \begin{matrix}
 &[F_1I_1]& [F_2I_1]&... &[F_1 I_2] & [F_2 I_2] & ... .\\
 [F_1I_1]&  \mathbf{A}_{11} \mathbf{A}_{11}  &   \mathbf{A}_{21}\mathbf{A}_{11} & ...\\
 [F_2I_1]&  & &  &   \mathbf{A}_{12}\mathbf{A}_{21} & \mathbf{A}_{22}\mathbf{A}_{21}  & ...\\
 ... \\
 [F_1I_2]&  \mathbf{A}_{11}\mathbf{A}_{12}  &   \mathbf{A}_{21} \mathbf{A}_{12} & ...\\
 [F_2I_2]&  & &  &   \mathbf{A}_{12}\mathbf{A}_{22}  & \mathbf{A}_{22}\mathbf{A}_{22} & ...\\
 ... \\
\end{matrix}
\end{equation}
In words, the entry of $\mathbf{M}$ corresponding to row $[F_i I_j]$ and column $[F_k I_l]$ equals one if there is a path $k \to l=j \to i$.
 
We observe that the nonzero eigenvalues of the matrix $\mathbf{M}$ are equal to those of the edge adjacency matrix $\mathbf{E}$. The rows and columns of $\mathbf{E}$ correspond to the existing edges of the network, and   $\mathbf{E}_{ij} = 1$ if edges $j$ and $i$ form a directed path. The eigenvalues are equal because of the symmetric structure of $\mathbf{M}$; if the $i$th column consists of zeros because $A_{kl}=0$, the same must be true for the $i$th row. Consequently, $\mathbf{M}$ corresponds to a linear transformation where the $i$th unit vector (corresponding to the $i$th column) collapses to the origo. As the $i$th component of all other columns is zero,  the transformation can be depicted in a lower-dimensional space (\emph{i.e.} the $i$th column and row can be removed) without affecting the nonzero eigenvalues. Finally, we note that the nonzero eigenvalues of the edge adjacency matrix are equal to the eigenvalues of the adjacency matrix $\mathbf{A}$. Consequently, $ \lambda_i(\mathbf{Y}) =  \beta \lambda_i(\mathbf{A}) -\beta-\delta$.

Given the eigenvalues $\lambda_i(\mathbf{Y})$, the eigenvalues of the matrix $\mathbf{C}$ can subsequently be found by setting each eigenvalue $\lambda_i(\mathbf{Y})$ to the diagonal of the middle block in $\mathbf{C}$ and solving the eigenvalues of this modified matrix $\mathbf{C_D}$. This is possible because the part of any eigenvector $\mathbf{e}(\mathbf{C})$ that $\mathbf{Y}$ modifies must be an eigenvector of $\mathbf{Y}$. To see this, we look at the eigenvalue equation  $\mathbf{C}\mathbf{e} = \lambda(\mathbf{C})\mathbf{e}$ where $\mathbf{e} = [\mathbf{e_1} \mathbf{e_2} \mathbf{e_3}]^T$. The multiplication results in equations
\begin{align}
-2\delta \mathbf{e_1} + \beta \mathbf{e_2} &= \lambda \mathbf{e_1} \label{eq21}\\
\mathbf{Y} \mathbf{e_2} + \gamma \mathbf{e_3} &= \lambda \mathbf{e_2} \label{eq22}\\
\delta \mathbf{e_1} + (-\gamma-\delta) \mathbf{e_3} &= \lambda \mathbf{e_3}. \label{eq23}
\end{align}
Substituting equations \ref{eq21} and \ref{eq23} into equation \ref{eq22}, we obtain
\begin{equation}
    \mathbf{Y} \mathbf{e_2} = \Big( \lambda(\mathbf{C}) + \frac{\beta\delta \gamma}{(\lambda(\mathbf{C})+\gamma + \delta)(-2\delta-\lambda(\mathbf{C}))}\Big) \mathbf{e_2}.
\end{equation}
As the first term on the right-hand side is a constant, it must be the case that $\mathbf{e_2}$ is an eigenvector of $\mathbf{Y}$.

The modified matrix $\mathbf{C_D}$ can be expressed as a Kronecker product 
\begin{equation}
\begin{bmatrix}
 -2\delta & \beta & 0 \\
 0 & \beta \lambda_i(\mathbf{A}) -\beta-\delta  &\gamma\\
\delta & 0 & -\gamma-\delta
\end{bmatrix}\mathbf{I}^{N^2 \times N^2}.
\end{equation}
Consequently,  the eigenvalues of $\mathbf{C}$ are given by the eigenvalues of the small $3 \times 3$ matrix above.
This matrix has the same form as the matrix in \cite{Droste}, and the signs of its eigenvalues can be solved similarly using the Hurwitz theorem. Solving for the critical value of $\lambda_1$, we obtain that the trivial steady state loses stability when
\begin{equation}
    \lambda_1 = \frac{\delta}{\beta} + \frac{\delta + \gamma/2}{\delta + \gamma}.
\end{equation}

\subsection{Critical drift in a Watts-Strogatz network} \label{appendix:ws_drift}

\begin{figure}[h!]
    \includegraphics[width=14cm]{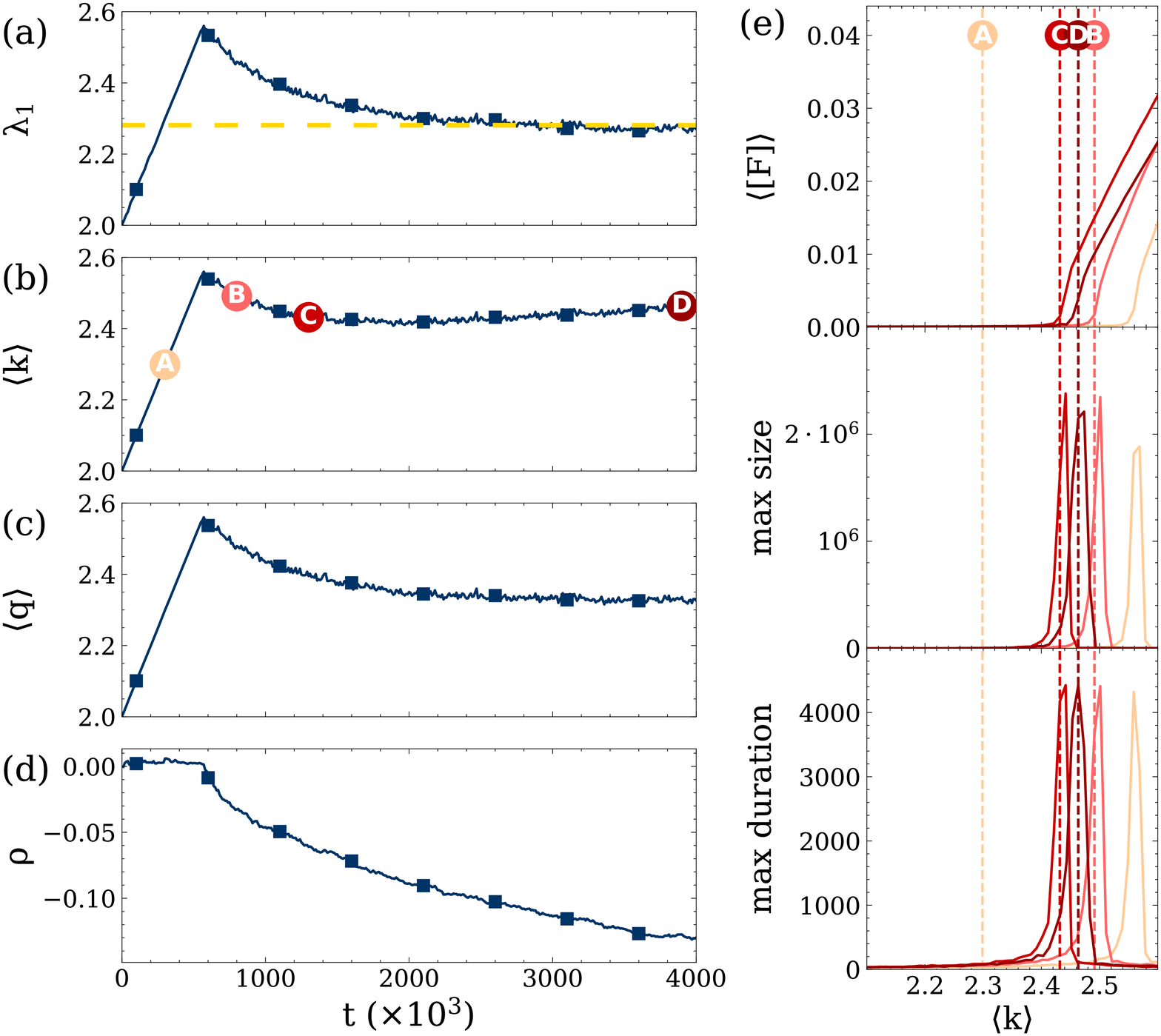}
    \caption{Drift in a network initialized as a directed Watts-Strogatz model. The graphs are obtained similarly to those in Fig. 2 in the main text, only the initial network structure differs.}
    \label{fig:ws_fig}
\end{figure}

As mentioned in the main text, the leading eigenvalue $\lambda_1$ of the network's adjacency matrix can change during the critical drift in some highly structured networks as well as in networks with a significant amount of loops. In such cases, the value of $\lambda_1$ alone does not determine whether or not the system resides at criticality. To illustrate this, we create a directed Watts-Strogatz network, a directed ring lattice where each node is connected to the next node as well as to the node directly after. When this network is evolved according to the adaptation rules, the  drift is clearly visible in all tracked parameters. Contrary to the initialization with ER networks, however, $\lambda_1$ does not stay constant during the drift [Fig.~\ref{fig:ws_fig}]. Instead, its value stabilizes only after the link removals and random additions have erased some of the original structure and increased randomness in the network.

\subsection{Correlation of the eigenvector centrality and the probability of firing} \label{appendix:eigc}

We show that a node's eigenvector centrality correlates with its probability of being in the firing state $F$. This has  previously been shown for several different epidemic models on undirected networks, including the SIS model (\cite{goltsev}). The proof for our directed IFRI model follows a similar pattern. Let us denote the probability of node $i$ being in state $F$ at time $t$ by $F_i$ and the probability of being in state $R$ by $R_i$. For simplicity, the corresponding vectors are denoted by $F$ and $R$. Assuming that the states of all nodes are independent of each other, the time evolution of the state vectors obey the differential equations
\begin{align}
    \dot F&= -\delta F + \beta I\circ  F\mathbf{A}   \label{eq1}\\
    \dot R&= \delta F - \gamma R, \label{eq2}
\end{align}
where $\mathbf{A}$ denotes the network's
adjacency matrix and $\circ$ denotes the Hadamard product. In addition, we know that
\begin{align}
    I &= 1 - F - R. \label{eq3}
\end{align}
At criticality, the derivatives equal zero as we are at a steady state. Substituting expressions \ref{eq2} and  \ref{eq3} into Eq. \ref{eq1}, we obtain
\begin{align}
    F&= \frac{\beta}{\delta}(1-F - \frac{\delta}{\gamma}F) \circ F \mathbf{A}
\end{align}
Assuming that the effect of terms involving $F\circ F$ is negligible, this can be expressed as
\begin{align}
    F &= \frac{\beta }{\delta}F\mathbf{A}.
\end{align}
For this equation to apply, it must be the case that $F$ is a left eigenvector of $\mathbf{A}$. 
If we assume that the network is strongly connected,  Perron-Frobenius theorem for non-negative matrices guarantees that the principal eigenvector is the only eigenvector where all elements are non-negative. Hence, as we require $F$ to be non-negative, it must correspond to the principal eigenvector, which gives the eigenvector centralities of individual nodes. If the network is not strongly connected, we can apply Perron-Frobenius theorem individually to each of the network's strongly connected components.

\subsection{Driving force of the drift}
\label{section:driving}

\begin{figure}[h!]
 \centering
\includegraphics[width=11cm]{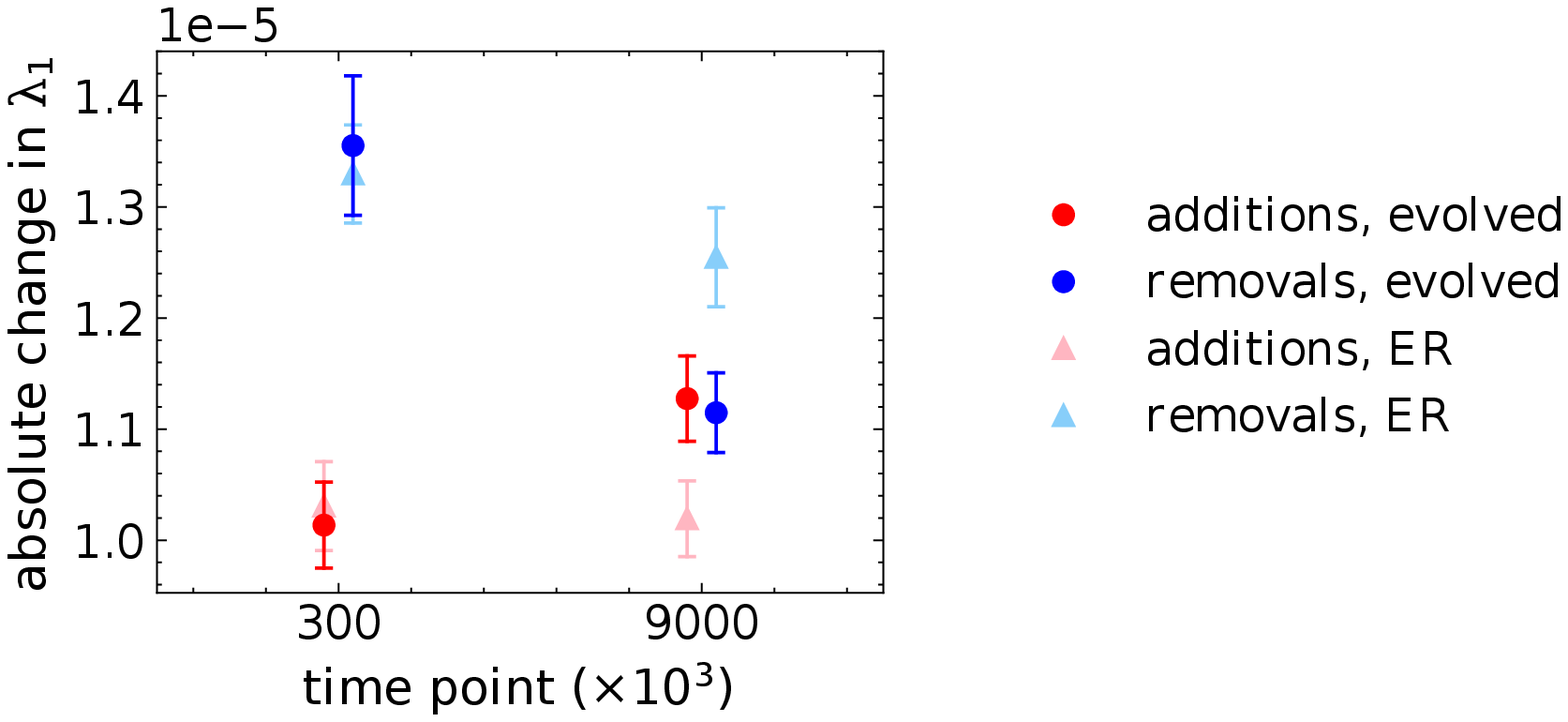}
    \caption{Average absolute change in the leading eigenvalue $\lambda_1$ after a random link addition (red circles) or targeted link removal (blue circles) in the evolved networks at two different times during the critical drift. 
    The triangles show the effect of a link addition/removal in static ER networks with mean degree equal to that of the evolved network at the corresponding time point. The markers show the mean effect of 10000 repeated link additions/removals with the bars displaying the 95\% confidence intervals. Initial $\langle k \rangle$ of the evolved network is 2, and the parameters are identical to those of Fig. 2 in the main text.}
    \label{fig:comparing}
\end{figure}

As discussed in the main text, the drift of the mean degree $\langle k \rangle$ results from the targeted link removals decreasing the leading eigenvalue $\lambda_1$ more efficiently than the random link additions increase it on average. We verify this by freezing a network at two different times during the critical drift, first at the beginning of the drift and second after the value of $\langle k \rangle$ has stabilized. In these frozen networks, we repeatedly remove or add a link to the network and measure the resulting change in $\lambda_1$ [Fig. \ref{fig:comparing}]. Since a node's eigenvector centrality correlates with its probability of being in the firing state, we simulate the targeted link removals from firing nodes by first choosing a node with probability proportional to its eigenvector centrality and then randomly removing one of its incoming links. 

We observe that at the beginning of the drift, the targeted link removals decrease $\lambda_1$ more efficiently than random link additions increase it on average, while at a later stage of the drift, the effects of link removal and addition have evened out. At this later stage, link removals have a weaker effect on $\lambda_1$ than in ER networks with the same $\langle k \rangle$, as expected. In addition, the network topology has changed in a way that the link additions increase $\lambda_1$ more than expected in an ER network with the same mean degree.

\newpage
\subsection{Dependence of the drift on parameters $\beta$, $\delta$ and $\gamma$}

It has been shown in \cite{Droste} that our model self-organizes to criticality irrespective of the specific choices of $\beta, \delta$ and $\gamma$ controlling the dynamics. In this section, we investigate how these parameters affect the critical drift of $\langle k \rangle$.

As argued in the previous section, the increase in $\langle k \rangle$ during the drift originates from the imbalanced effect of targeted link removals and random additions on the leading eigenvalue $\lambda_1$. If the network structure is such that the effects of a link addition and removal are already balanced when the system reaches criticality, we do not expect to observe this drift. 
Hence, as $\beta$, $\delta$ and $\gamma$ affect the critical value of $\lambda_1$ that the network self-organizes towards (see Eq. 1 in the main text), 
they indirectly influence the network topology at the beginning of the critical drift, which in turn determines whether or not $\langle k \rangle$ will increase during the drift.  

\begin{figure}[! h]
 \centering
    \includegraphics[width=10cm]{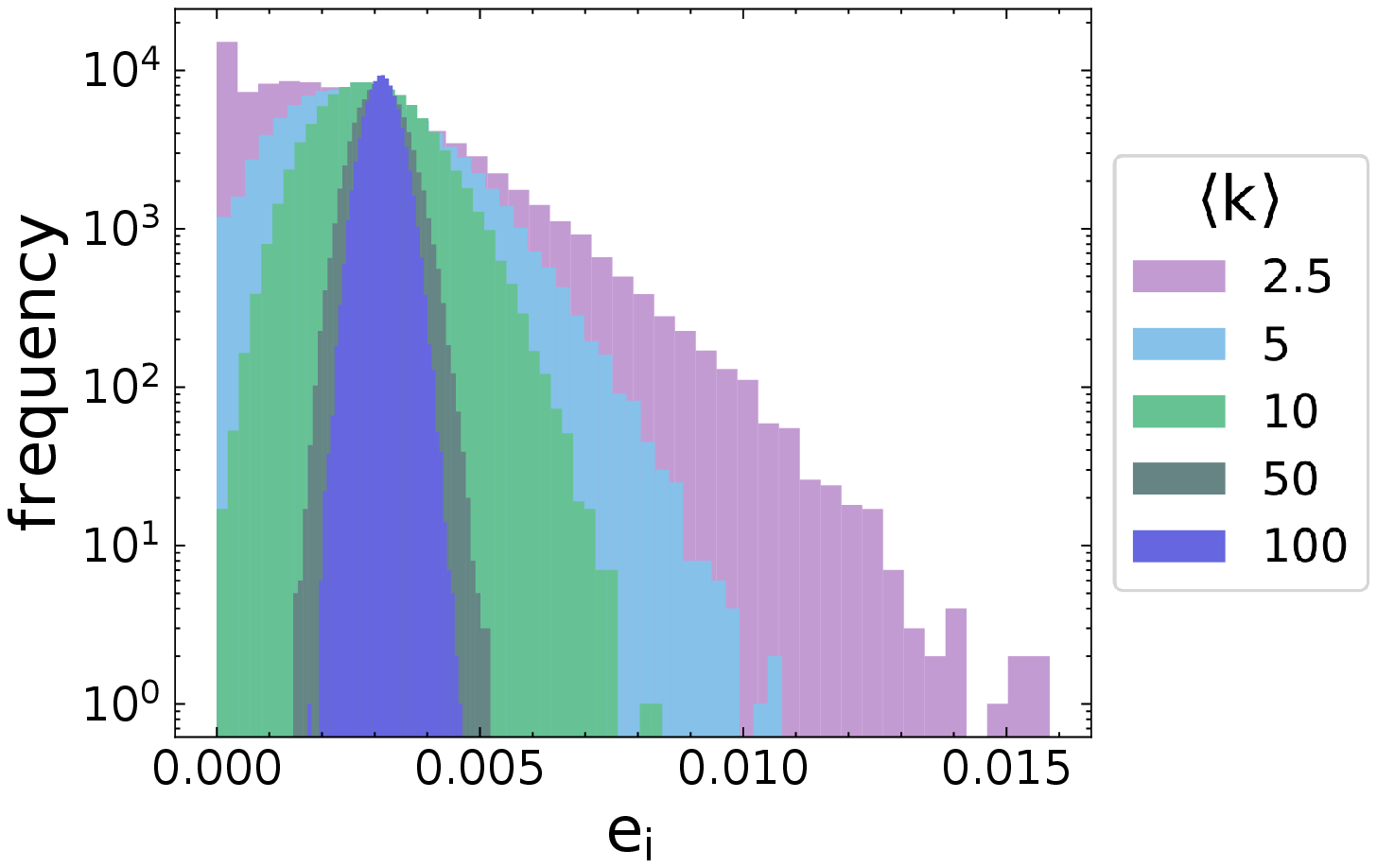}
    \caption{Histogram of the left principal eigenvector elements for static ER networks with different mean degrees. The size of the networks is $N=10^5$. The distributions of right principal eigenvector elements behave in a similar manner.} \label{fig:eig_dists}
\end{figure}

\begin{figure}[! h]
 \centering
    \includegraphics[width=7cm]{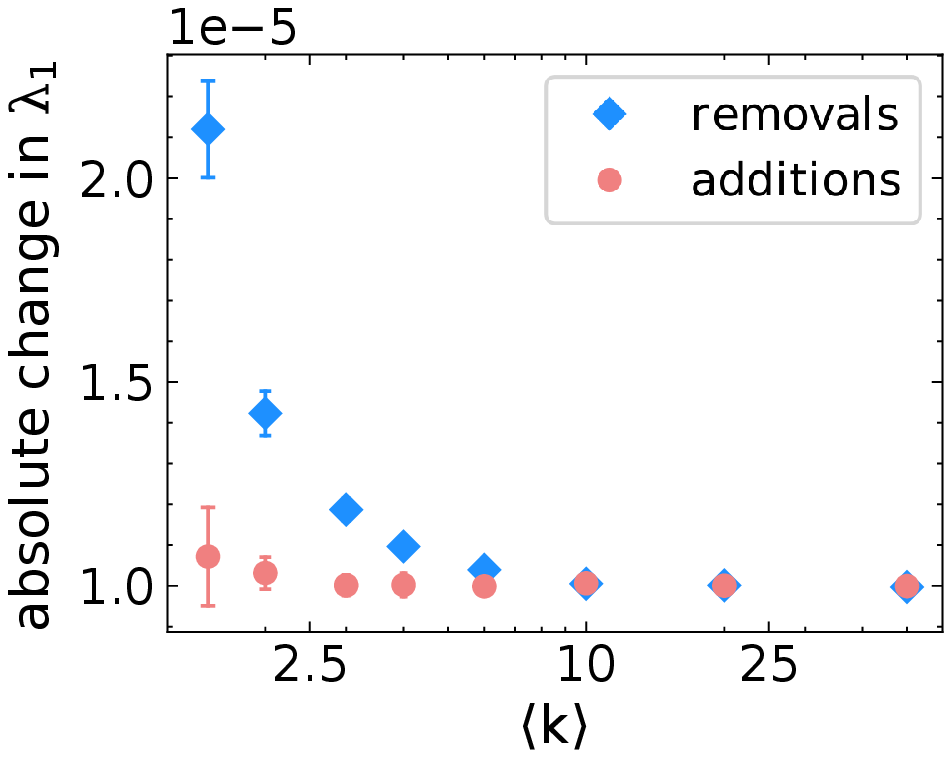}
    \caption{Change in $\lambda_1$ after a random link addition or targeted link removal for ER networks with different mean degrees. The targeted link removals are simulated by repeatedly choosing a node with probability proportional to its eigenvector centrality and randomly removing one of its incoming links. The markers show the mean of 10000 trials and largely overlay the 95\% confidence intervals.} \label{fig:er_effect}
\end{figure}

To understand the effect of parameters $\beta, \delta$ and $\gamma$ on the drift,  we first examine how the average effects of link addition and removal change when the mean degree of a static ER graph increases. As shown in Fig.~\ref{fig:eig_dists}, increasing $\langle k \rangle$ causes the distribution of eigenvector centralities (corresponding to the elements of the left principal eigenvector) to concentrate heavily around one value. This means that all nodes have an increasingly equal probability of firing, which causes the targeted link removals to reduce to random link removals. Consequently, the average value of a random link addition and a targeted removal converge to the same value, $1/N$ [Fig.~\ref{fig:er_effect}]. 

Since the drift of $\langle k \rangle$ is driven by the difference in the expected effect of a link removal and addition, we expect to observe this drift in ER networks only if the value of $\langle k \rangle$ is relatively low when the system first reaches criticality. This value can be approximated by the critical value $\lambda_1^*$ given by Eq. 1 in the main text (the approximation works especially well when the initial $\langle k \rangle$ at the start of the topological evolution is smaller than $\lambda_1^*$). As $\lambda_1^*$ is determined by the parameters 
$\beta, \gamma$ and $\delta$ controlling the dynamics, the combination of these parameters controls whether we observe the critical drift of $\langle k \rangle$ in ER networks.  Note, however, that this reasoning applies directly only to ER networks, and a network with a more varied topology and a more refined local structure may behave differently.

\begin{figure}[! h]
 \centering \includegraphics[width=15cm]{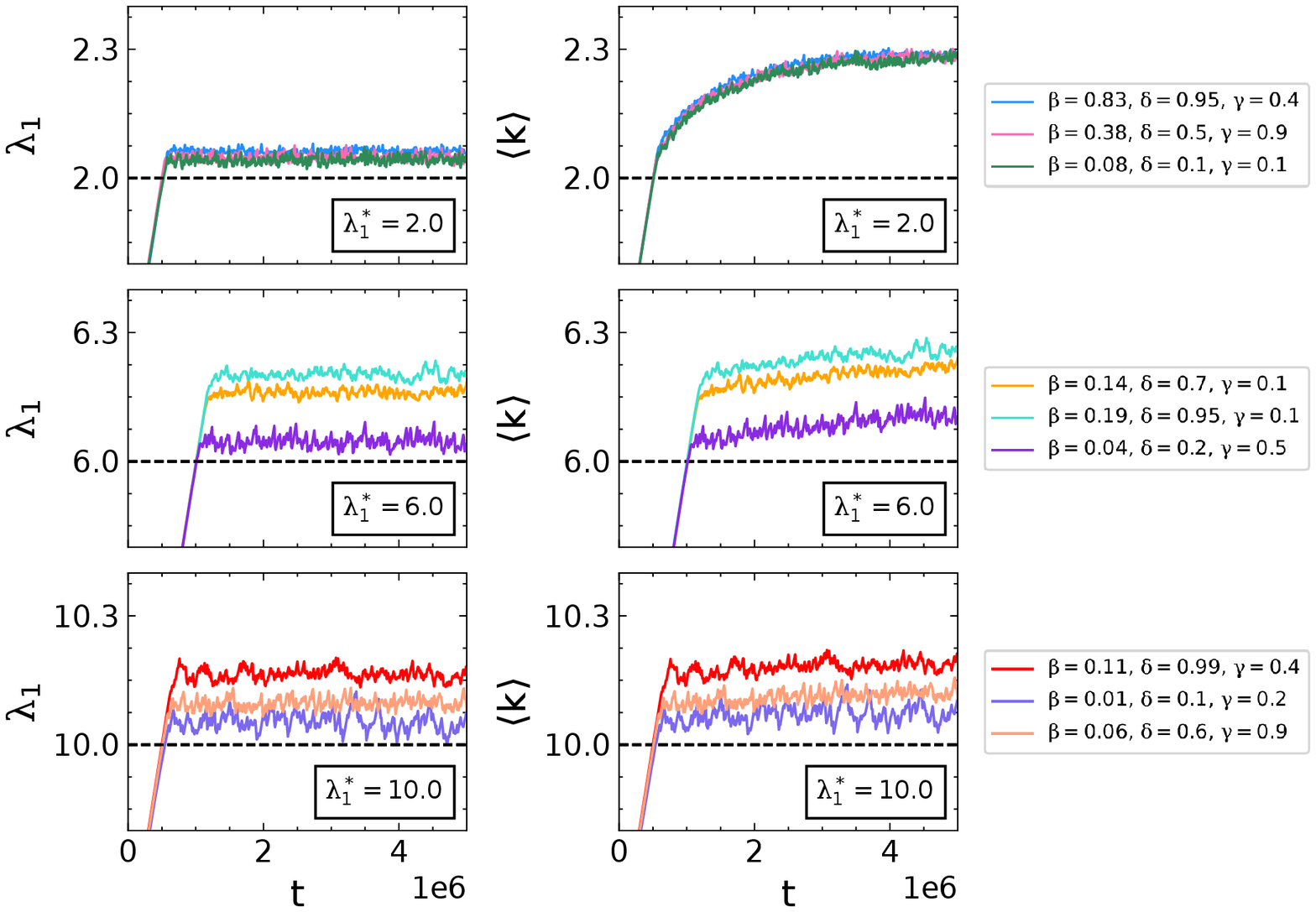}
    \caption{Time evolution of $\lambda_1$ and $\langle k \rangle$ for different combinations of parameters $\beta, \delta$ and $\gamma$. On each row, the parameter combinations have been chosen to produce a specific value of $\lambda_1^*$ (dashed lines) according to Eq. 1 of the main text. We have first chosen (with no specific logic) the values of $\delta$ and $\gamma$, after which the parameter $\beta$ is chosen so that Eq. 1 gives the desired value of $\lambda_1^*$. Other parameters are identical to those in Fig. 2 of the main text.} \label{fig:betadeltagamma}
\end{figure}

\begin{figure}[! h]
 \centering \includegraphics[width=15cm]{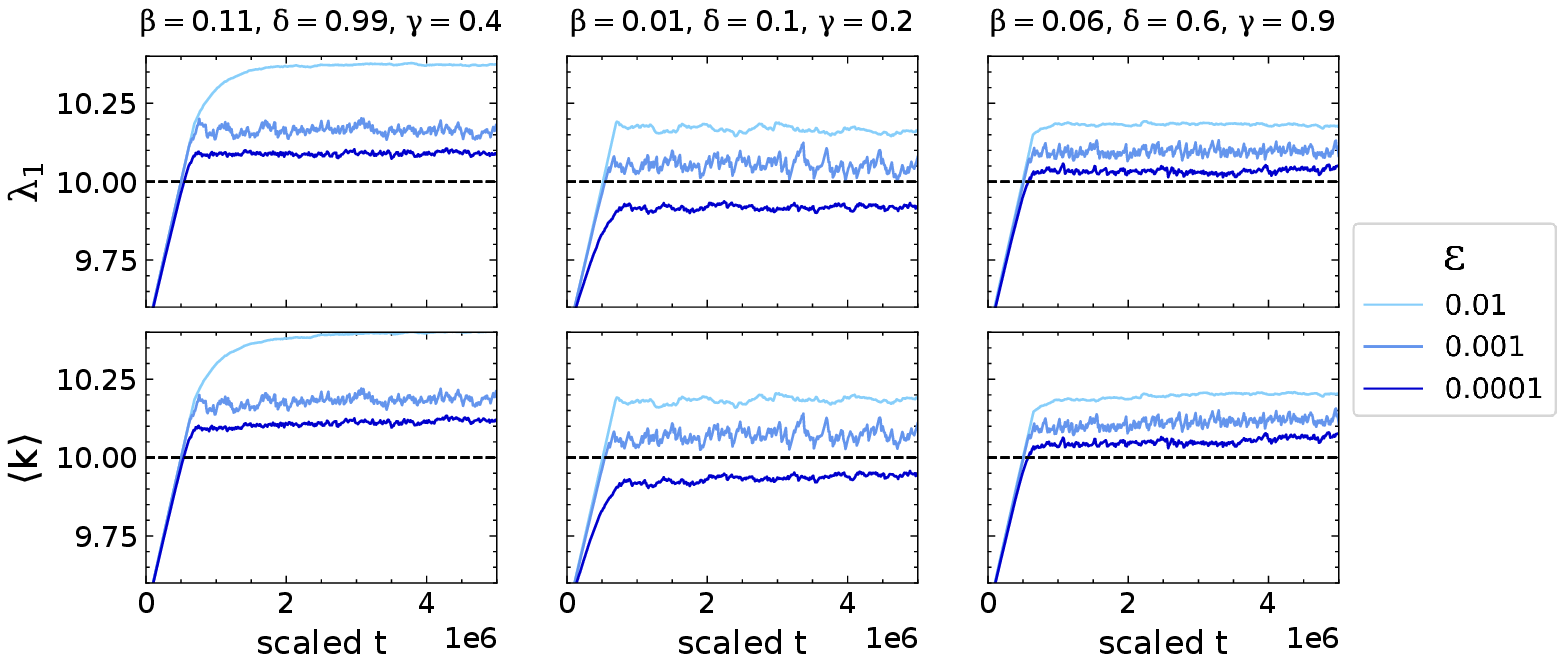}
    \caption{Time evolution of $\lambda_1$ and $\langle k \rangle$ for different values of $\epsilon$. Each column corresponds to one parameter combination resulting in $\lambda_1^*=10$ (dashed line) according to Eq. 1 of the main text.}
    \label{fig:betadeltagamma_epsilons}
\end{figure}

To verify the previous reasoning with simulations, we evolve networks with different combinations of parameters $\beta$, $\delta$ and $\gamma$ [Fig. \ref{fig:betadeltagamma}]. We observe that  the slope in which $\langle k \rangle$ increases during the drift seems to be unaffected by the specific parameter combination, and -- as expected -- this slope decreases as $\lambda_1^*$ increases. We also observe that
the parameter combination has an influence on how much the stabilized value of $\lambda_1$ deviates from the theoretical critical value $\lambda_1^*$. This latter observation, however, can be explained by the network evolution parameters $l$ and $\epsilon$ being finite. As already discussed in section \ref{appendix:parameters}b, the system self-organizes to criticality in the limit $l,\epsilon \to 0$, and for finite values of $l$ and $\epsilon$, the mean degree of the evolved networks is expected to deviate slightly from its critical value. \cite{Droste} have shown that for networks with a random ER structure, the magnitude of this deviation depends on $\delta, \gamma$ and $\langle k \rangle^*$ when $l$ and $\epsilon$ are finite. While this proof does not directly apply to our case, we expect to observe a similar dependency. Indeed, as shown in Fig. \ref{fig:betadeltagamma_epsilons}, decreasing the value of $\epsilon$ causes the evolved $\lambda_1$ to decrease towards its theoretical value or even below it, which 
 -- to our understanding -- is a result of the finite spontaneous firing rate pushing the system slightly towards the subcritical regime.

\subsection{Degree distribution of the evolved networks}

\begin{figure}[! h]
 \centering
    \includegraphics[width=14cm]{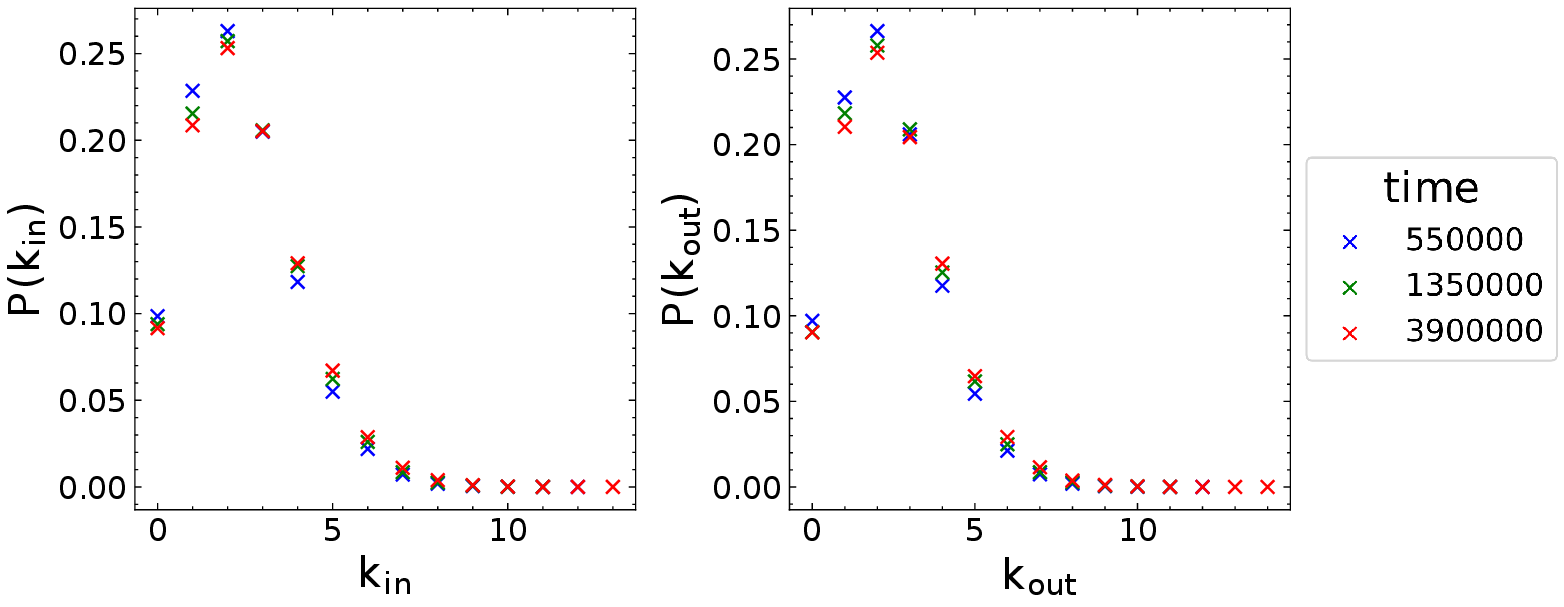}
    \caption{Distribution of in- and out-degrees at three different times during the drift}
    \label{fig:degdists}
\end{figure}

Figure \ref{fig:degdists} shows the distributions of in- and out-degrees at three different points during the drift.

\subsection{Time series of firing rates
}
\label{appendix:firing_rates}

\begin{figure}[! h]
 \centering
    \includegraphics[width=13cm]{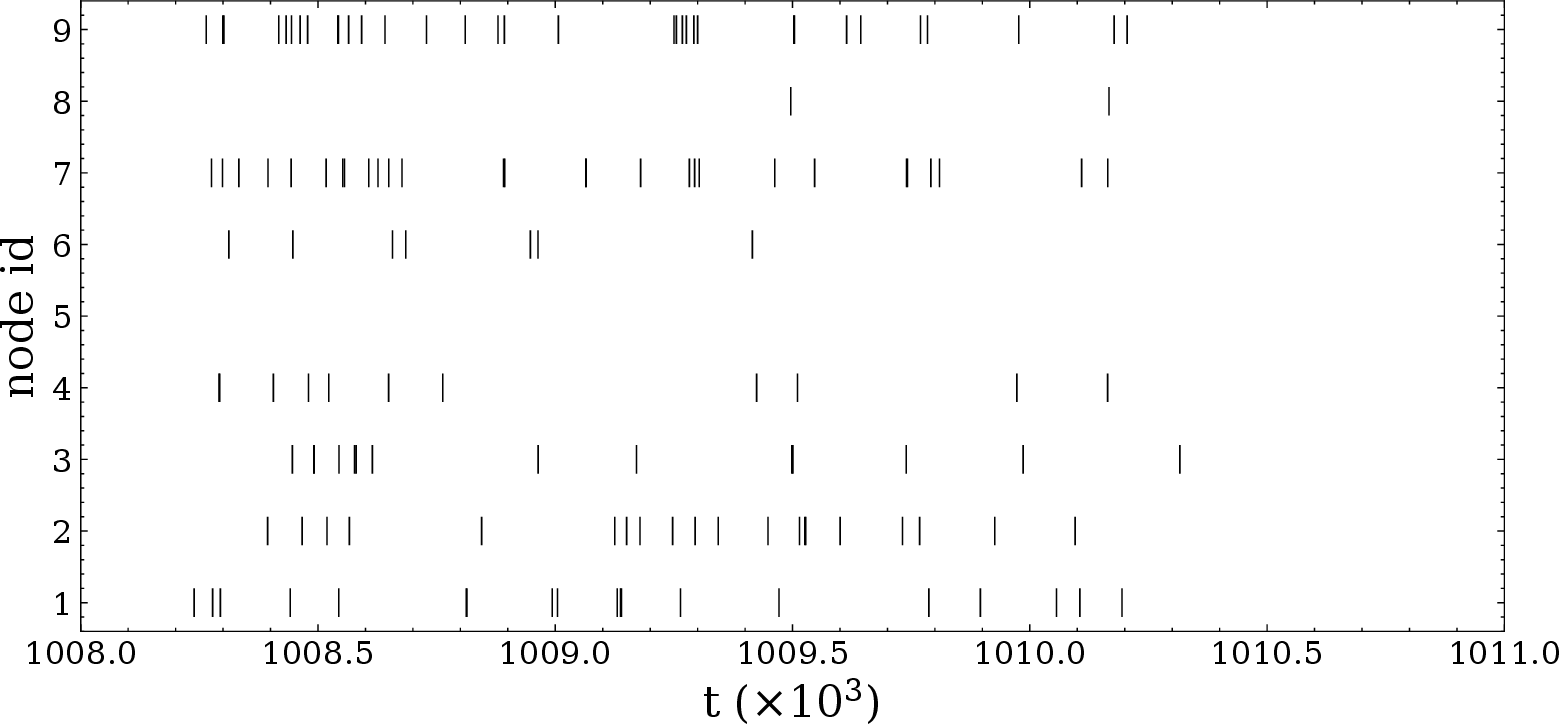}
    \caption{Firing rate time series of individual nodes. The parameters are identical to those in Fig. 2 of the main text.}
    \label{fig:firing_rates}
\end{figure}

Figure \ref{fig:firing_rates} shows a few examples of firing rate time series of individual nodes during a burst of activity along the critical drift. The nodes are chosen randomly. 

\clearpage

\providecommand{\noopsort}[1]{}\providecommand{\singleletter}[1]{#1}%
%
